%% file: mtaset.tex
\documentclass[conference]{IEEEtran}

\usepackage{cite}
\usepackage{amsmath,amssymb,amsfonts}
\usepackage{graphicx}
\usepackage{textcomp}
\usepackage{xcolor}

\usepackage{algpseudocode}
\usepackage{algorithm}
\usepackage{amsthm}
\usepackage{listings}
\usepackage{amsmath}
\usepackage{hyperref}
\newcommand{\arxiv}[1]{#1}

\arxiv{\usepackage{caption}
\usepackage{subcaption}}

\newtheorem{definition}{Definition}
\newtheorem{theorem}{Theorem}

\def\BibTeX{{\rm B\kern-.05em{\sc i\kern-.025em b}\kern-.08em
    T\kern-.1667em\lower.7ex\hbox{E}\kern-.125emX}}

\begin{document}

\title{MTASet: A Tree-based Set for Efficient Range Queries in Update-heavy Workloads}

\author{
    \IEEEauthorblockN{Daniel Manor}
    \IEEEauthorblockA{\textit{The Academic College of Tel Aviv-Yaffo}}
\and
    \IEEEauthorblockN{Mor Perry}
    \IEEEauthorblockA{\textit{The Academic College of Tel Aviv-Yaffo}}
\and
    \IEEEauthorblockN{Moshe Sulamy}
    \IEEEauthorblockA{\textit{The Academic College of Tel Aviv-Yaffo}}
}

\maketitle

\begin{abstract}
In concurrent data structures, the efficiency of set operations can vary significantly depending on the workload characteristics.
Numerous concurrent set implementations are optimized and fine-tuned to excel in scenarios characterized by predominant read operations. However, they often perform poorly when confronted with workloads that heavily prioritize updates. Additionally, current leading-edge concurrent sets optimized for update-heavy tasks typically lack efficiency in handling atomic range queries.
This study introduces the MTASet, which leverages a concurrent (a,b)-tree implementation. Engineered to accommodate update-heavy workloads and facilitate atomic range queries, MTASet surpasses existing counterparts optimized for tasks in range query operations by up to 2x. Notably, MTASet ensures linearizability.
\end{abstract}

\begin{IEEEkeywords}
Concurrent data structure, key-value map, set, dictionary, a-b tree, update-heavy, range query
\end{IEEEkeywords}


\section{Introduction}
\label{sec:intro}
Given the inherent challenges of concurrent programming, developers often use various concurrent data structures to build applications and complex systems, such as modern database engines designed for multicore hardware. These structures enable safe utilization in multithreaded environments through sophisticated synchronization algorithms optimized for performance.

The rise of multicore hardware has spurred the development of numerous new concurrent data structure designs, including dictionaries \cite{avni2013leaplist, basin2017kiwi, bronson2010practical, brown2012range, brown2011non, fomitchev2004lock, fraser2004practical, kobus2022jiffy} and sets \cite{braginsky2012lock, natarajan2014fast, shafiei2013non}.
These innovations consistently enhance performance over existing solutions and introduce features like atomic range scan operations (i.e., range queries).

Existing concurrent set or dictionary implementations typically excel in scenarios with low contention and predominantly read-oriented workloads, often neglecting the demands of update-intensive environments. Conversely, implementations optimized for update-heavy workloads frequently struggle with efficient range queries. 
For example, in experiments by Kobus et al. \cite{kobus2022jiffy}, SnapTree \cite{bronson2010practical} performs relatively well on update operations but exhibits poor performance on scan operations. Our research aims to address this issue by enhancing the scalability of range queries within a concurrent set optimized for update-heavy workloads, ensuring robust performance across diverse workload types.

Scaling range queries in concurrent data structures is inherently challenging due to the extensive coordination needed across elements.
Unlike single-key operations,
atomic range queries must traverse multiple nodes, risking inconsistent results if other threads update the data during the query,
thus adding further complexity to manage manage concurrent updates.
While locking or snapshotting methods can help maintain consistency, they impose synchronization overhead and lead to contention, especially in update-heavy environments. These issues make it difficult to efficiently support range queries at scale without compromising performance or data accuracy.

In response, we introduce MTASet, a concurrent set and dictionary with high update throughput that stores keys and their associated values and supports essential operations such as insertion, deletion, and lookup. In addition to high update throughput, MTASet is optimized for atomic range queries, which retrieve values for a specified range of keys.

MTASet uses a tailored multi-versioning approach \cite{bernstein1987concurrency} for atomic range queries, maintaining only the versions required for ongoing scans and managing version numbers through scans rather than updates. This significantly enhances the throughput of range query operations, especially under concurrent update-heavy workloads. Inspired by KiWi \cite{basin2017kiwi}, MTASet's range query demonstrates substantial performance gains in experimental evaluations, outperforming many state-of-the-art data structures in both read-mostly and update-heavy workloads.

MTASet is an (a,b)-tree, a variant of B-trees that allows between $a$ and $b$ keys per node, where $a \leq \frac{b}{2}$. It is based on a concurrent version of Larsen and Fagerberg's relaxed (a,b)-tree \cite{larsen1995b}, specifically the OCC-ABtree \cite{srivastava2022elimination}. MTASet employs fine-grained versioned locks to ensure atomic sub-operations and uses version-based validation in leaf nodes to guarantee correct searches. To manage overhead, MTASet incorporates established techniques, such as avoiding key sorting in leaves and minimizing unnecessary node copying.

The core philosophy of MTASet is to promptly handle client operations while deferring data structure optimizations to an occasional maintenance procedure.
This procedure, called \emph{rebalance},
aims to balance MTASet’s (a,b)-tree for faster access
and to eliminate obsolete keys through compaction.

\subsection{Background}
A \emph{set} data structure stores unique elements without any particular order. It supports three operations: \emph{Insert} adds an element to the set if it is not already in the set, \emph{delete} removes an element from the set if it exists, and \emph{contains}  tests whether an element is in the set.

An \emph{$(a,b)$-tree} \cite{black1998dictionary} is a balanced leaf-oriented search tree where each node can have between $a$ and $b$ children, and \mbox{$2 \leq a \leq \frac{b}{2}$}.
This tree structure optimizes operations by maintaining logarithmic height with respect to the number of elements, ensuring efficient data retrieval, insertion, and deletion.

The balanced structure of the \((a,b)\)-tree ensures logarithmic time complexity for all operations, making it suitable for applications requiring frequent insertions, deletions, and lookups, such as database indexing.

MTASet utilizes a \emph{concurrent} $(a,b)$-tree data structure, based on OCC-ABtree \cite{srivastava2022elimination}, which includes optimizations tailored for update-heavy workloads.
OCC-ABtree does not include a range query operation.
However, it was noted in \cite{srivastava2022elimination} that a range query capability could be implemented for the OCC-ABtree using 
the following technique detailed in \cite{arbel2018harnessing}:

In OCC-ABtree, leaf nodes are interconnected in a linked list, with each leaf node storing keys along with \emph{insertionTime} and \emph{deletionTime} fields indicating when keys were added and removed, respectively.
A global variable $TS$ is incremented atomically by a range query
from time $t$ to $t'$.
During insertion, $TS$ is read and written to the \emph{insertionTime} field
of the new key atomically.
During deletion $TS$ is read and written 
to the \emph{deletionTime} field of the deleted key,
and stored in the thread that executes the deletion
in a list of deleted keys accessible for other threads to read.
Special precautions are taken during deletion
to prevent race conditions.
A range query traverses leaf node lists, collecting keys with \emph{insertionTime} less than or equal to $t$.
It subsequently checks thread-specific lists
for keys deleted after time $t$,
using each key's \emph{deletionTime} to identify
missed deletions during traversal.

MTASet's range query introduces significant improvements,
detailed in Section~\ref{sec:evaluation},
that enhance the range query operation throughput
beyond this technique,
while maintaining the performance advantages
of the OCC-ABtree on update-heavy workloads.

Our correctness notion is \emph{linearizability}, which intuitively means
that the object "appears to be" executing sequentially.
It is defined for a \emph{history}, which is a sequence of operation
invoke and return steps, possibly by multiple threads.
A history partially orders operations: operation $op1$ \emph{precedes}
operation $op2$ in a history if $op1$'s return precedes $op2$'s invoke;
two operations that do not precede each other are \emph{concurrent}.
An object is specified using a \emph{sequential specification},
which is the set of its allowed sequential histories.
Roughly speaking, a history $\sigma$ is \emph{linearizable} \cite{lin_cor_cond}
if it has a sequential
permutation that preserves $\sigma$'s precedence relation and satisfies
the object's sequential specification.

\subsection{Contributions}
The primary contribution of this paper is the development and analysis of MTASet,
a concurrent set specifically optimized for high update throughput and frequent range queries.
MTASet enhances performance by leveraging a tailored multi-versioning approach,
which maintains only the necessary versions of keys for ongoing scans,
thus minimizing overhead and optimizing range query efficiency.
Notably, MTASet supports wait-free atomic range queries,
ensuring that range query operations complete in a bounded number of steps regardless of the workload.
By adapting concepts from OCC-ABtree and incorporating other techniques to manage versioning effectively,
MTASet significantly improves atomic range queries while sustaining high performance in update-heavy workloads.
Experimental results demonstrate that MTASet surpasses many existing concurrent data structures in both read-mostly and update-intensive scenarios.
This work addresses a critical gap in concurrent data structures, and can also be used as a robust framework for balancing update and range query operations,
making MTASet a valuable addition to the toolkit of multicore developers.

MTASet supports the following operations:
\begin{itemize}
     \item find($k$): Checks if a key-value pair with the key $k$ exists.
     If it does, the associated value is returned; otherwise, it returns $\perp$.
    \item insert($k$, $v$): Verifies if a key-value pair with the key $k$ exists.
    If it does, it returns the associated value; otherwise,
    it inserts the key-value pair and returns $\perp$.
    \item delete($k$): Deletes the key-value pair with the key $k$ if it exists
    and returns the associated value. Otherwise, it returns $\perp$.
    \item scan(fromKey, toKey): Returns the values of all keys within the range [fromKey, toKey].
\end{itemize}

\paragraph*{Evaluation results}
The MTASet Java implementation can be found on GitHub \cite{mtaset}. In Section~\ref{sec:evaluation}, we benchmark its performance under various workloads.
In most experiments, it significantly surpasses  OCC-ABtree*, a variant of OCC-ABtree\cite{srivastava2022elimination} tailored for update-heavy workloads with atomic range query capabilities \cite{arbel2018harnessing}. This positions MTASet as a concurrent set optimized for update-heavy tasks, offering efficient, atomic, and wait-free range queries.

The benefits of MTASet are evident in our primary scenario, which includes long scans amid concurrent update operations. MTASet did not outperform competitors \cite{basin2017kiwi} optimized solely for range scans but not for updates. However, in updates, MTASet significantly outperformed them, up to three times. In scenarios involving long scans with concurrent updates, MTASet exceeded the performance of the OCC-ABtree* \cite{arbel2018harnessing,srivastava2022elimination} by up to three times while maintaining comparable performance in update operations, thus preserving its update-heavy nature. Notably, MTASet's atomic scans are 1.6 times faster than the \emph{non-atomic} scans offered by the Java Skiplist written by Doug Lea \cite{lea2017concurrent} based on work by Fraser and Harris \cite{fraser2007concurrent}, and MTASet's updates are up to 3.6 times faster than those of the Java Skiplist.

\subsection{Related Work}

Various data structure designs and techniques have been developed to optimize performance in concurrent environments, focusing on skip lists, trees, and range query methods.

In the category of skip lists,
KiWi \cite{basin2017kiwi} is a Key-Value Map that supports linearizable,
wait-free range scans via a multi-versioned architecture similar to MTASet,
and its operations utilize the Compare-and-swap (CAS) atomic instruction for  lock-free functionality.
While KiWi achieves high throughput in range scans, it is not optimized for updates as MTASet is. LeapList \cite{avni2013leaplist} also supports linearizable range scans, employing fine-grained locks for concurrency control, similar to MTASet. Another related structure, Jiffy \cite{kobus2022jiffy}, is a linked-list data structure that offers arbitrary snapshots and atomic batch updates. Nitro \cite{lakshman2016nitro} leverages multiversioning to create snapshots, though these snapshots are not thread-safe during concurrent insert/remove operations.

In the category of tree-based structures, OCC-ABtree \cite{srivastava2022elimination} is a concurrent (a,b)-tree tailored for update-heavy workloads but lacks native range scan support. However, a general method for range queries is proposed \cite{arbel2018harnessing}, demonstrating lower throughput than MTASet. SnapTree \cite{bronson2010practical} is a lock-based, relaxed-balance AVL tree that provides atomic snapshots and range scans through a linearizable clone operation. Minuet \cite{sowell2012minuet} is a distributed, in-memory B-tree that enables linearizable snapshots using a costly copy-on-write approach, which allows snapshot sharing across multiple range scans. BCCO10 \cite{bronson2010practical} introduces a Binary Search Tree with optimistic concurrency control, similar to MTASet, utilizing version-based validation for efficient search operations. Additionally, LF-ABtree \cite{brown2017techniques} is a lock-free (a,b)-tree structure similar to the relaxed (a,b)-tree \cite{larsen1995b} employed in MTASet.

For range query techniques, Arbel-Raviv and Brown \cite{arbel2018harnessing} discuss implementing range queries in concurrent set data structures using epoch-based memory reclamation, proposing a traversal algorithm to ensure that all items within a range are accessed during the traversal’s lifetime. Nelson et al. \cite{nelson2022bundling} present a technique for achieving linearizable range queries on lock-based linked data structures.


\section{MTASet Algorithm}
\label{sec:algorithm}
In this section, we discuss the MTASet algorithm, exploring its core data structures, node types, and coordination mechanisms that support concurrent operations. We examine the specific roles of different node types, including leaf, internal, and tagged nodes, and the functionality of the linked-list structure of the leaf nodes. Furthermore, we discuss MTASet's operations, highlighting mechanisms such as versioning, locking, and the ongoing scans array (OSA). 

MTASet contains a permanent entry pointer to a sentinel node, a reliable starting point for all operations. This guarantees that every thread starts traversal from a well-defined, stable location.
The sentinel node contains no keys and a single child, the root node.

A node is \emph{underfull} if it contains fewer keys than the minimum $a$,
and it is \emph{full} when its number of keys equals the maximum $b$.

An example of the MTASet tree is shown in Fig.~\ref{fig:dstree}.

\subsection{Data structures}
MTASet has three types of nodes: leaf nodes, internal nodes and tagged internal nodes.

Each node contains a \emph{lock} field,
using MCS locks \cite{mellor1991algorithms}
where threads awaiting the lock spin on a local bit,  efficiently scaling across multiple NUMA nodes.
A thread modifies a node only if it holds the corresponding lock.
Leaf nodes include a \emph{version} field
which tracks the number of modifications made to the leaf and indicates whether it is currently changing.
Upon acquiring a lock, a thread increments the version before initiating modifications,
and increments it again
once it completed its changes before releasing the lock.
Thus, the version is even when the leaf is not being modified and odd when it is.
Searches utilize the version to ascertain whether any modifications occurred while reading the keys of a leaf.
Furthermore, nodes contain a \emph{marked} bit, toggled when a node is unlinked from the tree, allowing updates to determine whether a node is present in the tree.
Once marked, nodes are never unmarked.

MTASet's operations use a helper \emph{search} operation which returns a \emph{PathInfo} structure. It provides information about the node at which the search terminated, its parent and grandparent, the index of the node in the parent’s pointers array, and the parent index in the grandparent’s pointers array.

\subsubsection{Leaf Nodes}
\begin{figure}[htbp]
\centerline{\includegraphics[scale=0.83,]{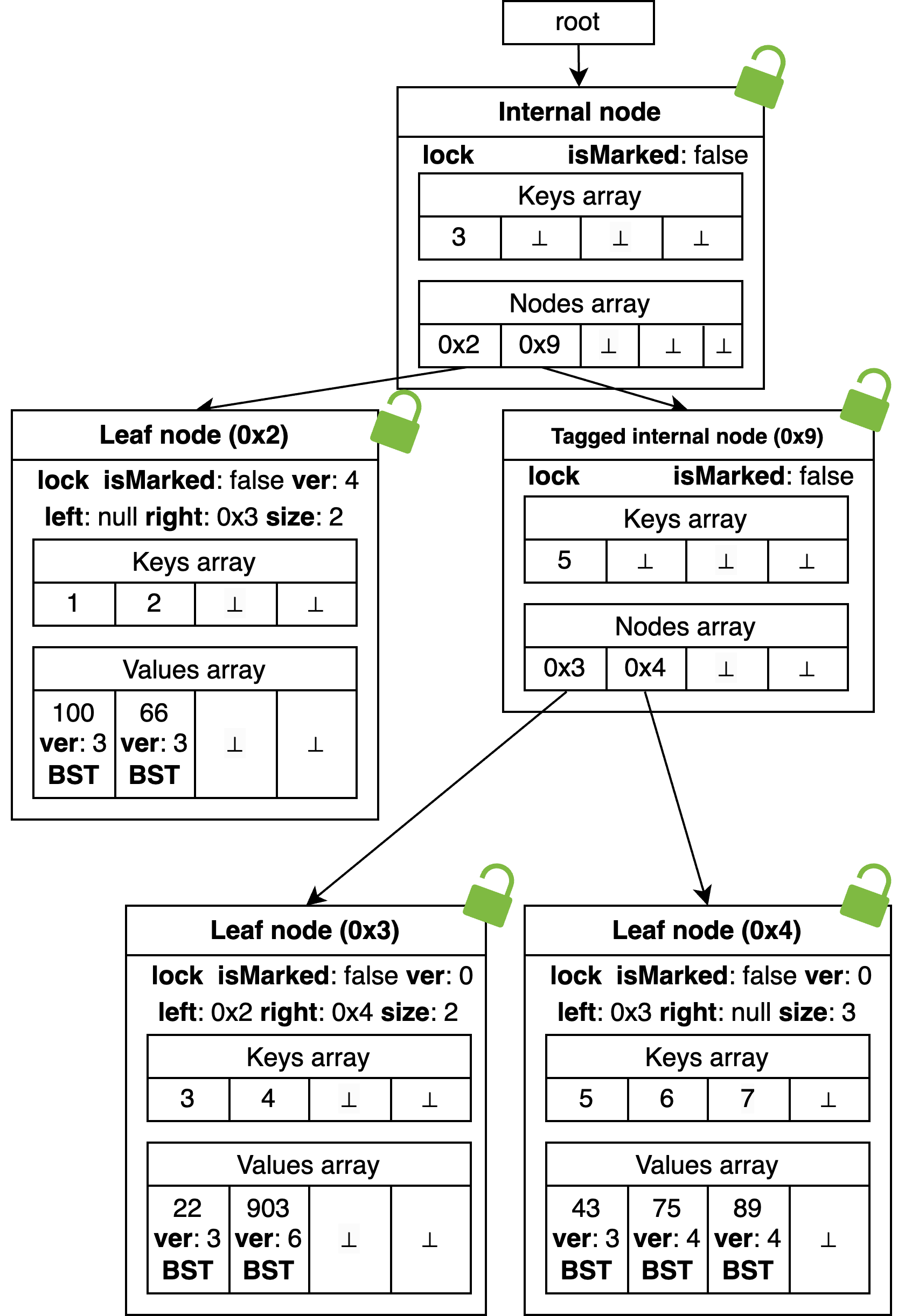}}
\caption{A snapshot of MTASet: An internal node pointing to a tagged internal node and a leaf node. The tagged internal node points to two leaf nodes. The locks are MCS, no locks are acquired}
\label{fig:dstree}
\end{figure}

Leaf nodes consist of arrays for keys and values.
A keys entry is considered empty when represented as $\bot$ and does not have a corresponding value,
as shown in Fig.~\ref{fig:dstree}.
The keys array is unordered, allowing for empty slots,
supporting faster updates by eliminating the need to rearrange keys during insertions and deletions.
The latest value and version of each key are kept in
the corresponding cell in the values array,
while older values are organized in a binary search tree.

Values are versioned, meaning they retain both the value for the latest version and values for past versions.
A value could be $\bot$, indicating a logical deletion in the corresponding version, or a non-$\bot$ value.

Neighboring leaf nodes are linked through left and right pointers.
This setup forms a linked list of leaf nodes with the property:
for each leaf $l$,
the keys in $l.right$ are strictly greater than those in $l$.
Rebalancing procedures, which involve linking and unlinking
leaf nodes due to occasional underfull or full conditions,
ensure that at any given time it is possible to reach
the right-most leaf node from the left-most leaf node.
This list aims to facilitate scan operations,
enabling it to traverse leaf nodes directly without traversing the entire tree,
as these are the only ones containing values.

\subsubsection{Internal Nodes}
An \emph{internal} node is a non-leaf node that serves as a routing point to direct searches through the structure towards the appropriate leaf nodes where data is stored. Internal nodes have two sorted arrays: one holding $k$ child pointers and the other holding $k - 1$ routing keys, which direct searches to the correct leaf. These routing keys remain constant. Adding or removing a key necessitates replacing the entire internal node, which occurs relatively infrequently. On the other hand, child pointers are mutable and subject to change.

A \emph{tagged} internal node is a non-leaf node that represents a height imbalance within the tree. It exists when a key/value insertion is required into a full node. Upon splitting the node, the two resulting halves are connected by a tagged node. Tagged nodes stand alone and are not involved in any other operations, consistently having precisely two children. They are eventually eliminated from the tree by invoking the fixTagged rebalancing step.

\subsubsection{Coordination data structure}
To coordinate scan and rebalancing operations,
MTASet utilizes a Global Version field (GV) and ongoing scans array (OSA),
which keeps track of the versions of ongoing scans
and is used by rebalancing for compaction purposes.
The OSA and GV are updated by the scan operation,
and emplyed by other operations and internal functions.
Their usage is described in the description of each operation in Section~\ref{operations}.
\arxiv{The full descriptions and algorithms of all operations are provided in Appendix~\ref{appendix:algorithm}
and the source code \cite{mtaset}}.

\subsection{Operations}
\label{operations}
Each operation invokes the Search function, which takes a key $k$ and traverses the tree from the root to locate the leaf node where $k$ resides.
This function is identical to the one used in OCC-ABtree \cite{srivastava2022elimination}.
The function searchLeaf locates a specified key \textit{k} within a leaf node \textit{l} and attempts to retrieve the corresponding value if \textit{k} exists in \textit{l}. Drawing inspiration from the classic double-collect snapshot algorithm \cite{afek1993atomic}, it executes as follows: Initially, it reads the version of the leaf \textit{l}. Then, it scans through \textit{l}'s keys array to locate \textit{k}. Afterward, it re-reads the version of \textit{l} to verify that no modifications occurred while retrieving the key and its associated value. If concurrent updates are detected, a retry is initiated. If no concurrent updates are detected and $k$ is found, searchLeaf returns \( (SUCCESS, \text{value}) \). If $k$ is not found, it returns \( (FAILURE, \perp) \).

Notably, both the search and searchLeaf functions are designed to run lock-free. This enhances concurrency by allowing updates to internal nodes to occur simultaneously with searches, boosting performance in environments with frequent reads and writes. 

The find($k$) operation is used by MTASet to locate the relevant leaf node and retrieve the value associated with $k$ in the tree. It simply calls the search and searchLeaf functions and returns the corresponding value.

\subsubsection{Insert and Delete}
During the insert(key, value) operation, a thread starts by executing a search(key, target) and searchLeaf(key,leaf) functions. The operation returns the associated value if the key is found during this search. Otherwise, it proceeds to lock the leaf and tries to insert the key (along with its corresponding value) into an available empty slot within the keys and values array. This process is known as a simple insert. However, if no empty slot is found, and considering that keys may become obsolete due to logical removals, the insert operation then checks for keys that can be physically removed (by invoking the cleanObsoleteKeys function, \arxiv{described in Appendix~\ref{appendix:algorithm}}). If obsolete keys are removed, the new key is inserted, and the fixUnderfull function is called to ensure the node meets the minimum size requirement. If the node's size falls below this minimum, it will either merge the underfilled node with a sibling or redistribute keys between them (using the fixUnderfull helper function).

If no obsolete keys are removed, the insert operation locks the leaf's parent and replaces the pointer to the leaf with a pointer to a newly created tagged node. This tagged node points to two new children: one containing the contents of the original leaf and the other containing the newly inserted key-value pair. This scenario is termed a splitting insert. The modification of the pointer, and thus the insertion of the key, occurs atomically. Following this, the insert operation invokes fixTagged \cite{srivastava2022elimination} to eliminate the tagged node from the tree.

Deleting a key involves writing (key, $\perp$) by calling the Insert function. If a key is not found or has already been logically deleted, $\perp$ is returned. If the key exists, the thread duplicates the current latest value into the key's version history data structure, sets the latest value with $\perp$, and then updates its version using CAS.

\subsubsection{Scan}
In the scan(lowKey, highKey) operation, a thread initially performs an atomic fetch-and-add operation on the GV (Global Version) global variable to increment its value. The obtained version is then published by writing it to the global, ongoing scan array (OSA). The thread also synchronizes with the rebalancing operation by atomically attempting to write the value read from GV using CAS.
Upon invoking the search operation, the thread identifies the node intended to contain lowKey. From this node, using the scanLeaf function, it traverses the leaf nodes, reading the values corresponding to keys within the [lowKey, highKey] range. These values meet the criteria of having a version equal to or less than the version in the OSA and are not $\perp$. Throughout this traversal, the thread ensures that the collected values are sorted by their keys in ascending order before being copied to the result array. The scan terminates by not proceeding to the next node upon encountering a key whose value exceeds highKey or upon reaching the end of the traversal path.
Finally, the scan information is removed from the OSA by writing $\perp$ to the appropriate cell. The operation then returns an array containing the scanned values along with its size.

\subsubsection{Helping updates}
The update operations (insert and delete) rely on the current value of GV, whereas a scan operation begins by atomically fetching and incrementing GV. This action ensures that all subsequent updates write versions greater than the fetched one. The scan then utilizes the fetched version, \textit{ver}, as its reference time, guaranteeing that it returns the latest version for each scanned key that does not surpass \textit{ver}. However, a potential race condition might arise if an update operation reads GV equal to \textit{ver} for its data and then pauses momentarily. Simultaneously, a concurrent scan fetches GV, equal to \textit{ver}, as its reference time. The scan may overlook or read the key before it is inserted or logically deleted with the version \textit{ver}. In this situation, the key should be included if inserted or excluded if it is deleted in the scan since its version equals the reference time, but it may not be due to its delayed occurrence. To tackle this issue, scans are designed to help updates by assigning versions to the keys they write. Concerning the update operations, they will write the key in the target node keys array without a version, read GV, and then attempt to set the version to the key's value using CAS. If a scan encounters a key without a version, it will attempt to help the update thread by setting the GV to the key version using CAS.

\subsubsection{Helping scans}
The cleanObsoleteKeys function, \arxiv{discussed in detail in Appendix~\ref{appendix:algorithm}}, is responsible for managing obsolete keys to support efficient memory use and consistency during compaction processes. 

A potential race condition may occur when a scan publishes its version on the OSA, and the cleanObsoleteKeys function requires a scan version for compaction purposes. This situation arises if a scan operation fetches (and increments) GV as its version and then pauses momentarily. Concurrently, the cleanObsoleteKeys function reads all current scan versions from the OSA and may overlook the scan version. Although the scan operation version should be utilized in this scenario, its delayed occurrence could prevent its consideration. Like scan operations helping updates (insert and delete), cleanObsoleteKeys is designed to help scans by assigning versions to them. Concerning the scan operation, it first publishes its data to the OSA without a version, fetches and increments GV, and then attempts to set the version to its published data using CAS. Suppose cleanObsoleteKeys encounters a published scan without a version. In that case, it will try to assist by fetching and incrementing GV and subsequently setting the fetched version to the published scan data using CAS. cleanObsoleteKeys will reread the scan’s version for its needs.


\section{Correctness}
\label{sec:correctness}

This section proves that MTASet is linearizable. To clarify, an algorithm achieves linearizability when, during any concurrent execution, each operation seems to occur atomically at a certain point between its invocation and its response. The linearizability of MTASet involves establishing a connection between the tangible representation of MTASet, the data stored in the system’s memory, and its conceptual set form. It involves demonstrating that the operations effectively modify the physical structure of the tree in a manner consistent with the abstract principles outlined at the end of Section~\ref{sec:intro}.

\subsection{Definitions}
\begin{definition}[Reachable Node]
\label{def_reachable_node}
A node is considered reachable if it can be accessed by traversing child pointers starting from the entry node.
\end{definition}

\begin{definition}[Key in MTASet]
\label{def_key_in_mtaset}
A key $k$ is in the tree if the following conditions are all met:
\begin{enumerate}
    \item It is in some reachable leaf $l$'s keys array.
    \item The version of $k$'s value in $l$ is set.
    \item The latest value of $k$ in $l$ is not $\perp$.
\end{enumerate}
\end{definition}

\begin{definition}[Key range]
\label{def_key_range}
The key range of a node is a half-open subset (e.g., [1,900)) of the set of all keys that can appear in the subtree rooted at that node.
\end{definition}

\begin{definition}[Node key range]
\label{def_node_key_range}
The key range of the entry node is the range of all keys present within the tree. Let $n$ be an internal node reachable with a key range of [L, R). If $n$ contains no keys, its child's key range remains as [L, R). However, if $n$ does contain keys $k_1$ through $k_m$, then the key range of $n$’s leftmost child (referred to by $n$.ptrs[0])
is [L, $k_1$), the key range of $n$’s rightmost child
(referred to by $n$.ptrs[m]) is [$k_m$, R). 
For any middle child referred by $n$.ptrs[$i$], the key range is [$k_i$, $k_{i+1})$.
Intuitively,
a node's key range represents the collection of keys permitted to exist within the subtree originating from that node.
\end{definition}

\begin{definition}[Search Tree]
\label{def_search_tree}
Let $n$ be an internal node within a tree, and let $k$ be a key within $n$.
A tree is a search tree when the following conditions are met:
\begin{enumerate}
    \item All keys within the subtrees to the left of $k$ in $n$ are strictly less than $k$.
    \item All keys within the subtrees to the right of $k$ in $n$ are either greater than or equal to $k$.
\end{enumerate}
\end{definition}

\subsubsection{Invariants}
We establish a set of invariants regarding the tree's structure. These invariants remain valid for the tree's initial state,
and any alteration to the tree upholds all of these invariants.
These established invariants are a foundation for proving the data structure's linearizability.

\begin{theorem} MTASet Invariants: The following invariants are true at every configuration in any execution of MTASet:

\begin{enumerate}
    \item All reachable nodes (Definition \ref{def_reachable_node}) form a relaxed (a,b)-tree. \label{invar_relxedab}
    \item The key range (Definition \ref{def_key_range}) of a reachable node that was removed remains constant. \label{invar_keys_in_deleted_node}
    \item An unreachable node (which does not satisfy Definition \ref{def_reachable_node})
    retains the same keys and values it held when it was last reachable and unlocked, meaning updates do not simultaneously detach and alter a node. \label{invar_unreached_node_retains}
    \item Each key appears only once in a leaf node among all leaf nodes. \label{invar_Key_appear_once}
    \item If a node was once reachable and is presently unmarked, it remains reachable. \label{invar_reachble_is_unmarked}
    \item Let $l1$ be a full or underfull leaf node that is part of a merge or split operation, and let $l2$ be a new node created by the split or merge. The leaf node  $l1$ can not reach $l2$ using $l1.right$ pointer.
    \label{invar_full_underfull_no_duplicate}
    \item Let $l$ be a linked node that is about to be unlinked, then $l.right$ and $l.left$ are constant (may never change once it is unlinked) \label{invar_leaf_node_adj}
    \item In the search operation on a node with key $k$ and target $t$, the key range of $n$ contains $k$. \label{invar_search_key_range_contains_key}
\end{enumerate}

\end{theorem}
Intuitively, invariants \ref{invar_relxedab} through \ref{invar_Key_appear_once} stem from the sequential accuracy of the updates, alongside the assurance that any node subject to replacement or modification remains locked and accessible until the update takes effect. The correctness of the updates in a single-threaded execution can be discerned through examination of the pseudocode, thus we refrain from a detailed proof. A brief clarification of invariants \ref{invar_relxedab} through \ref{invar_Key_appear_once} regarding concurrent correctness can be found in \cite{srivastava2022elimination}. Invariants \ref{invar_reachble_is_unmarked}, \ref{invar_full_underfull_no_duplicate}, and \ref{invar_leaf_node_adj} can be straightforwardly deduced from the pseudocode. Invariant \ref{invar_search_key_range_contains_key} differs slightly as it focuses on verifying the correctness of an operation rather than a structural property.
A detailed proof can be found in \cite{srivastava2022elimination}.

\subsubsection{Linearizability of Find}
The linearizability of the find operation is established by ensuring that the result accurately reflects the tree’s state at a specific moment during the search interval. During a find operation, if the target key $k$ is found, or if its absence is confirmed by either reading a placeholder value \( \perp \) or by scanning the entire key array within a leaf node $l$ that was unlocked during the search interval, then the result represents an accurate and stable view of $l$’s state at that moment.

If $l$ was part of the tree at any point during the unlocked interval of the search, then the result of the find operation can be linearized to any instant within this interval when $l$ was verified as part of the tree.
However, if $l$ was not part of the tree during the unlocked interval,
then its absence implies that it was unlinked from the tree concurrently with the search.
In this case, the find operation is linearized to the instant just before $l$ was unlinked, ensuring that each node visited in the traversal was part of the tree at some point during the operation.

Therefore, whether the leaf $l$ was present or unlinked, the result of find accurately reflects the tree’s state in real time at one distinct instant during the search interval, meeting the requirements for linearizability.

\subsubsection{Linearizability of Insert and Delete}
The linearizability of the insert operation in MTASet involves four potential linearization points. First, if the insert operation \( i \) finds its target key \( k \) during the search, it linearizes similarly to the find operation, returning the value corresponding to the located key. Second, if the insert operation \( i \) finds the target key \( k \) after acquiring the lock on leaf \( l \), it can linearize at any point while holding \( l \)’s lock. During this time, the key \( k \) and value cannot change, and the leaf \( l \) cannot be unlinked, ensuring the correctness of the returned value. Third, when \( i \) inserts into a non-full leaf \( l \) or modifies a logically deleted key, the operation linearizes at the moment the key \( k \) and value are written to their respective arrays with a version. Before this point, the key \( k \) did not exist in MTASet, afterward, it is part of MTASet with a non-\( \perp \) value. For splitting inserts, linearization occurs when the new subtree pointer is written to the parent node. Before this, the key \( k \) is not in MTASet, afterward, it is in one of the newly created leaf nodes, with other keys correctly assigned to new nodes.

For a full explanation, \arxiv{refer to Appendix~\ref{correctness:insert}}.
The linearization of the delete operation and the justification of return values follow a similar rationale as the first three cases of insert linearizability.

\subsection{Linearizability of Scan}
The linearizability of a scan operation \( s \) is established by the moment when the global version \( GV \) surpasses the version \( v_s \) used by \( s \) to collect values, ensuring that any key inserted or deleted after this point is excluded from \( s \)’s result. Despite ongoing rebalancing operations that dynamically link and unlink leaf nodes, these changes do not compromise the correctness of \( s \).

Specifically, when \( s \), linearized at time \( t \), initiates from the designated leaf node containing the smallest key in its specified range, it maintains correctness through several cases: if \( s \) encounters a node \( n \) that is concurrently unlinked, it proceeds directly to the next node, avoiding revisits and thereby preserving a consistent snapshot. In cases where \( s \) visits new nodes that replaced underfull nodes, any missing keys in \( n \) were removed prior to time \( t \) and thus are not included in \( s \). Similarly, if \( s \) encounters nodes created by a split of a full node, the key that triggered the split was inserted after \( t \) and therefore will not be collected by \( s \).

For a full explanation, \arxiv{refer to Appendix~\ref{correctness:scan}}.


\section{Evaluation}
\label{sec:evaluation}

In this section, we compare MTASet with the OCC-ABtree* \cite{srivastava2022elimination} implemented with range scan \cite{arbel2018harnessing},
OCC-ABtree without a range query implementation,
KiWi \cite{basin2017kiwi},
and Java's ConcurrentSkipList (non-atomic) \cite{lea2017concurrent}.

\subsection{System and setup}
In our experiments, we utilized a virtual machine on Azure (Standard\_D96ads\_v5) with the following specifications: an AMD EPYC 7763 64-Core Processor with 96 vCPUs and 384 GB of RAM. All data structures were implemented in Java. Both MTASet and the OCC-ABtree were configured with a=2 and b=256. The machine was running Ubuntu 20.04.2 LTS.

\subsection{Methodology}
Each experiment begins with a seeding phase,
where a random subset of integer keys and values are inserted into
the data structure until its size reaches half of the key range.
Following this, 80 threads are created and started simultaneously, consisting of \( k \) threads designated for scans and \( 80-k \) threads for other operations, marking the start of the measured phase of the experiment. During this phase, each of the \( 80-k \) threads repeatedly selects an operation (insert, delete, find) based on the desired frequency of each operation. This phase lasts 10 seconds, recording the total throughput (operations completed per experiment).
Threads designated for the scan operation repeatedly perform scans, recording the total number of collected keys. Each experiment is conducted 10 times, and our graphs display the averages of these runs.

\subsection{Experiments}
\begin{figure}[htbp]
    \centerline{\includegraphics[scale=0.78]{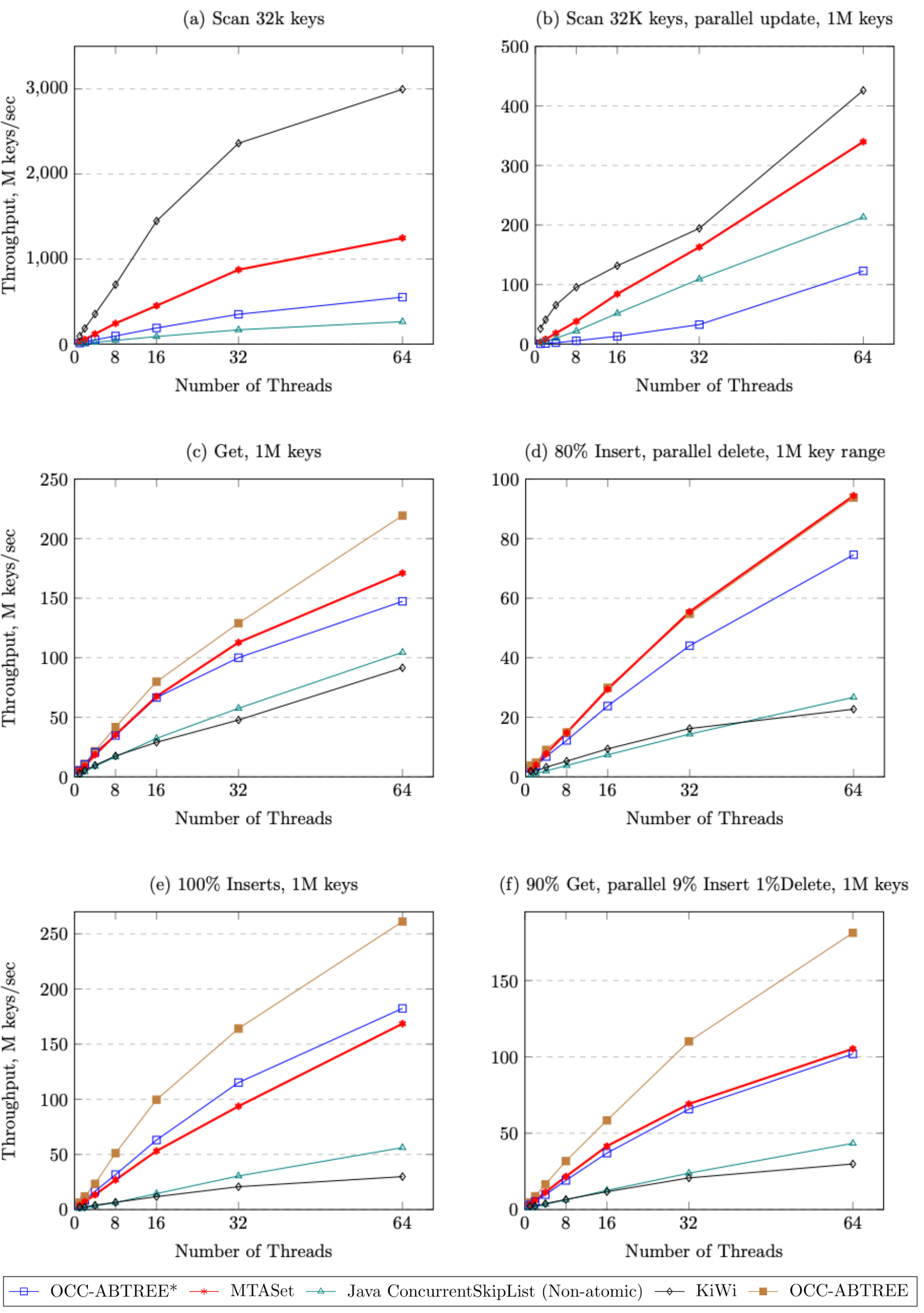}}
    \caption{Experiments (a) and (b) measure Scan throughput, (c) and (f) measure Get throughput, and (d) and (e) measure Insert throughput.
    (*) OCC-ABtree implemented with scan support.}
    \label{fig:benchmark}
\end{figure}

We present six experiments (a)-(f), with results shown in Fig.~\ref{fig:benchmark}.
In the 100\% scan experiment (a), all threads perform only range queries. KiWi achieves the highest throughput, which is expected as it is optimized for range scans, whereas MTASet is designed for intensive updates. However, the gap narrows in the scan-with-parallel-updates experiment (b), where threads perform range queries in parallel with updates. Here, MTASet’s scan throughput significantly surpasses OCC-ABtree* by nearly fivefold, highlighting the impact of this comparison.

In the get experiment (c), where all threads only perform find operations, OCC-ABtree takes the lead, and MTASet performs slightly better than OCC-ABtree*. In experiment (d), with 80\% inserts alongside deletes, MTASet and OCC-ABtree (without range query support) achieve the highest throughput, with MTASet’s throughput nearly four times that of KiWi. In the 100\% inserts experiment (e), OCC-ABtree leads, showing the best performance, with OCC-ABtree* slightly outperforming MTASet.

In the 90\% get, 9\% insert, and 1\% delete experiment (f), where threads perform 90\% Get in parallel with 9\% Insert and 1\% Delete, MTASet performs comparably to OCC-ABtree* and far exceeds both KiWi and the non-atomic JavaConcurrentSkipList. 

Overall, MTASet significantly outperforms OCC-ABtree* in range scan workloads while maintaining competitive update performance. This balance makes MTASet a strong candidate for workloads that require efficient range queries without sacrificing significant update efficiency.

\section{Discussion}
\label{sec:discussion}
In this study, we introduced MTASet, a concurrent set data structure designed to excel in both high update throughput and efficient, wait-free atomic range queries. MTASet combines a multi-versioning approach to optimize range query performance while preserving exceptional efficiency in environments with intensive updates. Notably, MTASet demonstrates a significant advantage over the OCC-ABtree*  in range query operations, while maintaining competitive performance in update-heavy workloads. The results from our experiments show that MTASet strikes a remarkable balance between these two operations, making it a versatile solution for modern applications requiring both fast updates and efficient range scans.

Looking ahead, an exciting avenue for future work is the potential integration of elimination \cite{eliminationStack,eliminationArray,exchanger} into MTASet. This enhancement could further improve its range query performance, offering additional optimizations for diverse workload types and solidifying MTASet as a robust, balanced data structure for concurrent applications.


\bibliographystyle{IEEEtran}
\bibliography{mtaset}

\arxiv{
\clearpage
\appendix
\input{appendix}
}

\end{document}

%% file: appendix.tex
\section{MTASet Algorithm}
\label{appendix:algorithm}
In this appendix, we present the MTASet algorithm, detailing its core and helper functions. Figures and pseudocode are included for clarity. We cover the find, insert, delete, and scan operations, including specific handling for logically deleted keys. Additionally, we describe the methods for rebalancing and maintaining key-value consistency through versioning.

\subsection{Insert Key}
The insertKey (Algorithm \ref{alg:insertKey}) function accepts a key, a value, and a leaf node as inputs. It writes the given key and value to the specified leaf node and attempts to atomically assign a version to the inserted value. If the leaf node is full, the function returns RETRY, signaling to the caller that the insertion failed. If the insertion is successful, it returns the inserted value.

\subsection{Update Key}
The updateKeyInIndex function (Algorithm \ref{alg:updateKey}) accepts an index, a value, and a leaf node as inputs. It writes the value to the specified leaf node at the given index and attempts to atomically assign a version to the inserted value. This function is used when the caller knows the exact cell for insertion. It is used when a key is logically deleted, but not physically, and a new key-value pair needs to be inserted.

\subsection{Can Update Key}
The canUpdateKeyInIndex function (Algorithm \ref{alg:canUpdateKey}) takes a key, value, and leaf node as inputs and determines whether it is possible to update the corresponding value for the key. Updating the key is possible only if the key is logically deleted or if the key exists and needs to be (logically) deleted. The function returns (true, i) if the key can be updated at index i, (false, i) if it cannot be updated at index i, and (false, $\perp$) if the key does not exist.

\subsection{Create Tagged Internal Node}
The createTaggedInternalNode function (Algorithm \ref{alg:createTaggedInternal}) takes a key, value, leaf, the index of the leaf in its parent node's array, and the parent node as inputs. It creates a tagged internal node that points to two new leaf nodes, which evenly distribute the given key along with all the keys from the specified leaf node. Additionally, it connects the new leaf nodes to the linked list of leaf nodes.

\begin{algorithm}
	\caption{Insert Operation} 
	\begin{algorithmic}[1]
 \tiny	
\Function{insert}{key, value}
    \While {True}
        \State path = \Call{search}{key, NULL}
        \State retCode, retValue = \Call{searchLeaf}{path.node, key}
        \If {retCode = SUCCESS \textbf{and} value $\neq \perp$ \textbf{or} retCode = FAILURE \textbf{and} value = $\perp$}
            \State \Return retValue
        \EndIf
        \State leaf, parent = path.node, path.parent
        \State \Call{Lock}{leaf}
        \If {leaf.marked}
            \State Unlock leaf
            \State \textbf{Continue}
        \EndIf
        \State canUpdateKeyInIndexResult $\gets$ \Call{canUpdateKeyInIndex}{key, value, leaf}
        \If{canUpdateKeyInIndexResult.Result}
        \State updateKeyInIndexResult $\gets$ \Call{updateKeyInIndex}{canUpdateKeyInIndexResult.Index, value, node}
        \State Unlock leaf
        \State \Return updateKeyInIndexResult
        \ElsIf {canUpdateKeyInIndexResult.Index $\neq$ $\perp$}
        \State Unlock leaf
        \State \Return $\perp$
        \EndIf
        \State currSize $\gets$ leaf.size
    \If{currSize $<$ this.maxNodeSize}
        \State result $\gets$ \Call{insertKey}{key, value, leaf}
        \State Unlock leaf
        \State \Return result
    \Else
           \State removedKeys $\gets$ \Call{cleanObsoleteKeys}{leaf}
    \If{removedKeys $>$ 0}
        \State writeResult $\gets$ \Call{insertKey}{key, value, leaf}
        \State \Call{unlock}{leaf}
        \State \Call{fixUnderfull}{leaf}
        \State \Return writeResult
    \EndIf
    \State Lock parent
    \If{leaf.left and/or leaf.right parent is not equal to leaf's parent}
    \State lock and check if marked. If marked, \textbf{Unlock} 
    leaf.left and/or leaf.right, leaf, parent and \textbf{Continue}
    \EndIf
    \State newTaggedInternal $\gets$ \Call{createTaggedInternalNode}{key, value, leaf, path.nodeIndex, parent}
    \State Unlock leaf and parent
    \State fixTagged(newTaggedInternal)
    \State \Return $\perp$
    \EndIf
    \EndWhile
\EndFunction
\end{algorithmic} 
\label{alg:insert}
\end{algorithm}

\begin{algorithm}
	\caption{Search} 
	\begin{algorithmic}[1]
\Function{Search}{key, targetNode}
    \State gp $\gets \perp$, p $\gets \perp$, pIdx $\gets$ 0, n $\gets$ entry, nIdx $\gets$ 0
    \While{$n$ is not Leaf}
        \If{$n$ = targetNode}
            \State \textbf{break}
        \EndIf
        \State gp $\gets$ p, p $\gets$ n, pIdx $\gets$ nIdx
        \State n $\gets$ entry, nIdx $\gets$ 0
        \While{nIdx $<$ node.size - 1 \textbf{AND}
        \newline key $\geq$ node.keys[nIdx]}
            \State nIdx++
        \EndWhile
        \State n $\gets$ n.ptrs[nIdx]
    \EndWhile
    \State \textbf{return} PathInfo(gp, p, pIdx, n, nIdx)
\EndFunction
\end{algorithmic} 
\label{alg:search}
\end{algorithm}

\begin{algorithm}
	\caption{Insert key}
	\begin{algorithmic}[1]
        \Function{insertKey}{key, value, node $n$}
    \For{$i \gets 0$ \textbf{to} $\text{this.maxNodeSize}-1$}
        \If{\text{n.keys}[i] = $\perp$}
            \State $\text{n.version} \gets \text{n.version} + 1$
            \State $\text{vc} \gets \text{new ValueCell}(key)$
            \State $\text{vc.putNewValue}(value)$
            \State $\text{n.values}[i] \gets \text{vc}$
            \State $\text{n.keys}[i] \gets \text{vc.key}$
            \State $\text{n.values}[i].\text{casLatestVer}(0, \text{GlobalVersion}())$
            \State $\text{n.size} \gets \text{n.size} + 1$
            \State $\text{n.version} \gets \text{n.version} + 1$
            
            \State \Return \text{node.values}[i].\text{latestValue}
        \EndIf
    \EndFor
    \State \Return \text{RETRY}
\EndFunction
\end{algorithmic} 
\label{alg:insertKey}
\end{algorithm}

\begin{algorithm}
	\caption{Update key} 
	\begin{algorithmic}[1]
\Function{updateKeyInIndex}{keyIndex, value, node}
    \State $\text{node.version} \gets \text{node.version} + 1$
    \State $\text{vc} \gets \text{node.values[keyIndex]}$
    \State $\text{vc.putNewValue(value)}$
    \State $\text{vc.casLatestVer(0, GlobalVersion)}$
    \State $\text{node.version} \gets \text{node.version} + 1$
    \State \Return \text{vc}.\text{latestValue}
\EndFunction
\end{algorithmic} 
\label{alg:updateKey}
\end{algorithm}

\begin{algorithm}
\caption{Can Update Key in Index} 
\begin{algorithmic}[1]	
\Function{CanUpdateKeyInIndex}{key, value, leaf}
        \For{$i \gets 0$ \textbf{to} $\text{this.maxNodeSize}-1$}
            \If {leaf.keys[i] = key}
                \If {leaf.values[i].latestValue $\neq \perp$ \textbf{and} value = $\perp$ \textbf{or} leaf.values[i].latestValue = $\perp$ \textbf{and} value $\neq \perp$}         
                \State \Return (true, i)
                \Else \State \Return (false, i)
                \EndIf       
            \EndIf
        \EndFor
        \State \Return (false, $\perp$)
\EndFunction
\end{algorithmic} 
\label{alg:canUpdateKey}
\end{algorithm}

\begin{algorithm}
\caption{Create Tagged Internal Node} 
\begin{algorithmic}[1]
\Function{createTaggedInternalNode}{key, value,
\newline leaf, leafIdxInParent ,parent}
    \State N $\gets$ contents of leaf $\cup$ $\{key/value\}$
    \State initialize two leaf nodes, $child1$ and $child2$
    \State newTaggedInternal $\gets$ TaggedInternal with two children: child1 and child2, who evenly shares $N$
    \State child1.right $\gets$ child2
    \State child1.left $\gets$ leaf.left
    \State child2.left $\gets$ child1
    \State child2.right $\gets$ leaf.right
    \State leaf.right.left $\gets$ child2
    \State leaf.left.right $\gets$ child1
    \State parent.ptrs[leafIdxInParent] $\gets$ newTaggedInternal
    \State leaf.marked $\gets$ true
    \State \Return newTaggedInternal
\EndFunction
\end{algorithmic} 
\label{alg:createTaggedInternal}
\end{algorithm}

\begin{algorithm}
\caption{Search Leaf}
\begin{algorithmic}[1]
\small 
\Function{searchLeaf}{leaf, key}
    \While{True}
        \State ver1 $\gets$ leaf.ver
        \If{ver1 is odd}
            \State \textbf{continue}
        \EndIf

        \State value $\gets \perp$
        \For{$i \gets 0$ \textbf{to} NodeMaxSize - 1}
            \If{leaf.keys[i] = key}
                \State value $\gets$ leaf.values[i].latestValue
                \State \textbf{break}
            \EndIf
        \EndFor
        \State ver2 $\gets$ leaf.ver

        \If{ver1 $\neq$ ver2}
            \State \textbf{continue}
        \EndIf
        \If{value = $\perp$}
            \State \textbf{return} (FAILURE, $\perp$)
        \Else
            \State \textbf{return} (SUCCESS, value)
        \EndIf
       
    \EndWhile
\EndFunction
\end{algorithmic}
\label{alg:searchLeaf}
\end{algorithm}

\begin{algorithm}
\caption{Find}
\begin{algorithmic}[1]
    \Procedure{find}{$key$}
        \State pathInfo $\gets$ PathInfo()
        \State searchResult $\gets$ \textsc{search}(key, \text{null}, pathInfo)
        
        \If{searchResult.\text{getReturnCode()} $\neq$
        \newline\text{ReturnCode.SUCCESS}}
            \State \textbf{return} Result(ReturnCode.FAILURE)
        \EndIf
        
        \State leaf $\gets$ pathInfo.n
        \State searchLeafResult $\gets$ \textsc{searchLeaf}(leaf, key)
        
        \State \textbf{return} searchLeafResult
    \EndProcedure
\end{algorithmic}
\label{alg:find}
\end{algorithm}

\subsubsection{Rebalancing}

\paragraph*{FixTagged} removes a tagged node from the tree. It is identical to the one used in the OCC-ABTREE. For pseudo-code and further explanation, see \cite{srivastava2022elimination}.

\begin{figure}[H]
    \centering
    \includegraphics[scale=0.6]{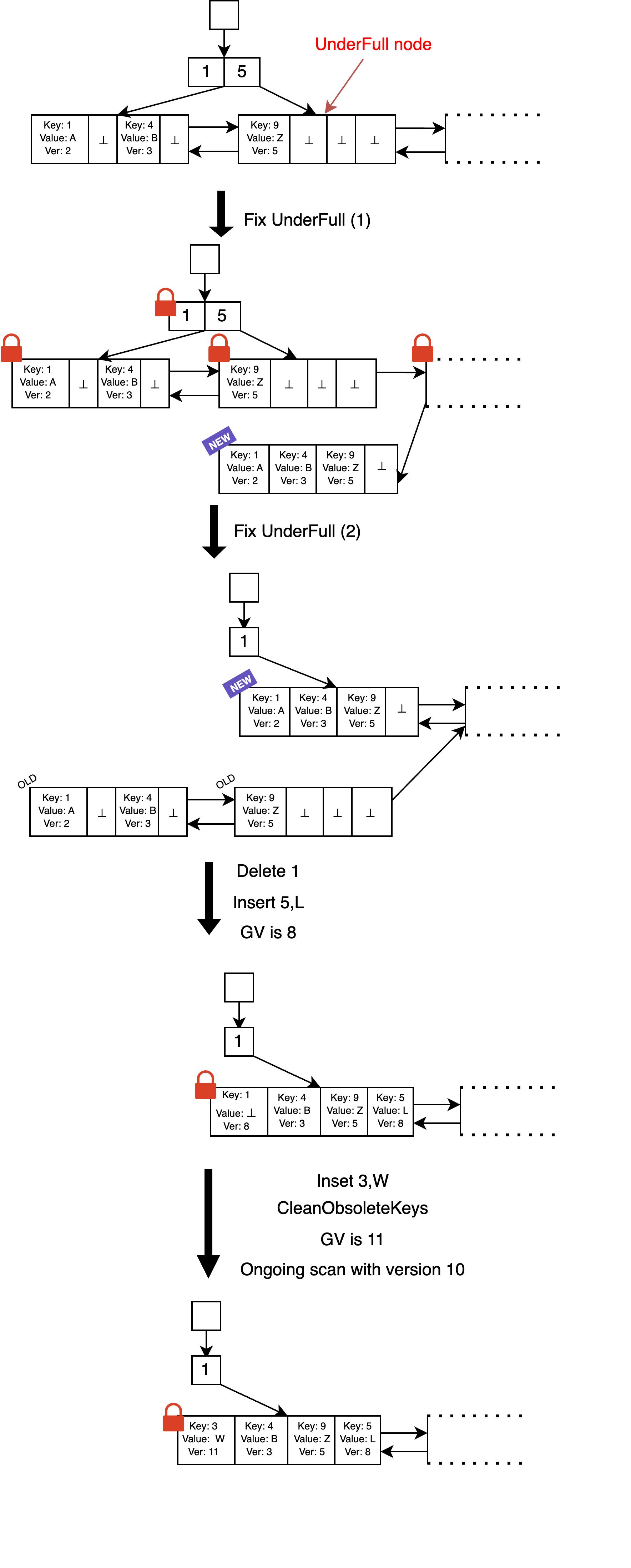}
    \caption{An underfull node is being merged with its sibling, and CleanObsoleteKeys removes obsolete keys}
    \label{fig:mtasetop}
\end{figure}

\paragraph*{FixUnderfull} (Algorithm \ref{alg:fixUnderFull}) addresses a node $n$ that falls below the minimum size unless $n$ is the root or entry node. It accomplishes this by evenly distributing keys between $n$ and its sibling $s$, provided that this action does not result in either of the new nodes becoming underfull. Alternatively, if redistribution is not feasible, fixUnderfull merges $n$ with $s$. In this scenario, the merged node may still be underfull, or the parent node might become underfull if it was already at the minimum size before merging its children. Consequently, fixUnderfull is recursively called on both the merged node and its parent. It is crucial for fixUnderfull that $n$ is underfull, its parent $p$ is not underfull, and none of $n$, $p$, and $s$ are tagged, and if $n$ is a leaf node, n's adjacent leaf nodes must be unmarked. If these conditions aren't met, fixUnderfull retries its search. (see Figure \ref{fig:mtasetop} for a visual representation of this process)

\paragraph*{CleanObsoleteKeys} (Algorithm \ref{alg:cleanObsoleteKeys}) removes keys logically deleted from a leaf node, creating space for new keys and reducing the need for rebalancing. Initially, the process involves identifying the smallest version of the current scan, $minScanVersion$, by iterating through the ongoing scan array (OSA) and retrieving the version of each ongoing scan. Subsequently, the function loops over the keys array of the leaf node to identify any deleted keys (where the latest version of the corresponding value is $\perp$). If a deleted key is found, and its latest version is less than or equal to $minScanVersion$, it is physically removed from the node, replacing it with $\perp$.  (see Figure \ref{fig:mtasetop} for a visual representation of this process)

\begin{algorithm}
\caption{Clean Obsolete Keys}
\begin{algorithmic}[1]
\small 
\Function{cleanObsoleteKeys}{node}
    \State minVersion $\gets \infty$
    \State numberOfCleanedKeys $\gets 0$
    \For{$i \gets 0$ \textbf{to} \text{ongoingScansArraySize} - 1}
        \State scanData $\gets$ \text{OngoingScansArray}[i]
        \If{scanData = $\perp$}
            \State \textbf{continue}
        \EndIf
        \State scanDataVersion $\gets$ scanData.version
        \If{scanDataVersion = 0}
            \State newVersion $\gets$ \text{GV.getAndIncrement()}
            \If{CAS(scanData.version, 0, newVersion)}
                \State scanDataVersion $\gets$ newVersion
            \Else
                \State scanDataVersion $\gets$ scanData.version
            \EndIf
        \EndIf
        \If{scanDataVersion $<$ minVersion}
            \State $minVersion \gets scanDataVersion$
        \EndIf
    \EndFor

     \For{$i \gets 0$ \textbf{to} \text{MAXNODESIZE} - 1}
        \State valueCell $\gets$ node.values[i]
        \If{valueCell = $\perp$}
        \State continue
        \EndIf
        \State latestValue $\gets$ valueCell.helpAndGetValueByVersion($\infty$)
        \If{latestValue $\neq \perp$}
            \State \textbf{continue}
        \EndIf
        \State 
        \State latestVersion $\gets$ valueCell.getLatestVersion()
        \If{minVersion $\geq$ latestVersion}
            \State node.version++
            \State node.keys[i] $\gets 0$
            \State node.values[i] $\gets \perp$
            \State node.size--
            \State node.version++
            \State numberOfCleanedKeys++
        \EndIf
    \EndFor
    \State \textbf{return} numberOfCleanedKeys
\EndFunction
\end{algorithmic}
\label{alg:cleanObsoleteKeys}
\end{algorithm}

\subsubsection{ScanLeaf}
The Scan Leaf function (Algorithm \ref{alg:scanLeaf}) takes a leaf node, a version, and a Scan operation range defined by low and high values as inputs. It scans through the leaf's key array to collect keys within the [low, high] range that have a value with a version less than or equal to the specified version. If there are multiple versions meeting this criterion, it selects the largest one. The function returns an array containing the collected keys and their corresponding values. Additionally, it provides a flag indicating whether any keys exceeding the high value were encountered during the scan.

\begin{algorithm}
\caption{Scan Leaf}
\begin{algorithmic}[1]
\tiny 
\Function{scanLeaf}{leaf, version, low, high}
    \State kvArray $\gets$ \textbf{new} KeyValue[maxNodeSize]
        \State kvArraySize $\gets 0$
        \State continueToNextNode $\gets$ \textbf{true}
        \For{$i \gets 0$ \textbf{to} maxNodeSize - 1}
            \State key $\gets$ leftNode.keys[i]
            \If{key = $\perp$}
                \State \textbf{continue}
            \EndIf
            \State valueCell $\gets$ leftNode.values[i]
            \If{valueCell = $\perp$}
                \State \textbf{continue}
            \EndIf

            \If{key $\geq$ low \textbf{AND} key $\leq$ high}
                \State value $\gets$ valueCell.helpAndGetValueByVersion(myVer)
                \If{value = $\perp$}
                    \State \textbf{continue}
                \EndIf
                \State kvArray[kvArraySize] $\gets$ (key, value)
                \State kvArraySize++
            \EndIf
            \If{key $>$ high}
                \State continueToNextNode $\gets$ \textbf{false}
            \EndIf
        \EndFor
        \State \Return (kvArray, kvArraySize, continueToNextNode)
\EndFunction
\end{algorithmic}
\label{alg:scanLeaf}
\end{algorithm}

\subsubsection{Create New Version}
The createNewVersion function (Algorithm \ref{alg:newVersion}), utilized by the Scan (Algorithm \ref{alg:scan}) operation, reads and increments the Global Version. It then creates and stores an object in the OSA at the cell index matching the executing thread ID. This object holds the retrieved version, indicating that a scan, linearized at the read version, is currently running.

\begin{algorithm}
\caption{Create New Version}
\begin{algorithmic}[1]
\Function{NewVersion}{scanData}
    \State \Call{publishScan}{scanData} \Comment{insert scanData to shared array (OSA)}
    \State myVer $\gets$ GV.getAndIncrement()

    \If{CAS(scanData.version,0, myVer)} \Comment{Compare-and-Swap}
        \State \textbf{return} myVer
    \Else \Comment{version was already set by a different thread}
        \State helpedVer $\gets$ scanData.version
        \State \textbf{return} helpedVer
    \EndIf
\EndFunction
\end{algorithmic}
\label{alg:newVersion}
\end{algorithm}

\subsubsection{Help and Get Value by Version}
The helpAndGetValueByVersion function (Algorithm \ref{alg:helpAndGetValueByVersion}) accepts a ValueCell and a version as inputs. It returns the value with a version equal to or less than the specified version. If it encounters a value without a version, it attempts to assign one by reading from the global version and using a CAS (Compare-And-Swap) operation.

\begin{algorithm}
\caption{HelpAndGetValueByVersion}
\begin{algorithmic}[1]
\small 
\Function{helpAndGetValueByVersion}{valueCell, version}
    \If{valueCell.version = 0}
        \State CAS(0, valueCell.version, GV)
    \EndIf
    \State vv $\gets$ valueCell.version 
    \If{vv $\leq$ version}
        \State \Return valueCell.value
    \EndIf
    \State \Return \text{this.previousVersions.floor}(version)
\EndFunction
\end{algorithmic}
\label{alg:helpAndGetValueByVersion}
\end{algorithm}

\begin{algorithm}
\caption{Scan}
\begin{algorithmic}[1]
\tiny 
\Function{Scan}{low, high, entry, result}
    \State myVer $\gets$ \Call{newVersion}{low, high} 
    \State resultSize $\gets$ 0
    \State pathInfo $\gets$ search(low, $\perp$)
    \State leftNode $\gets$ pathInfo.n
    \While{true}
        \State scanLeafResult $\gets$ \Call{scanLeaf}{leftNode, myVer}  
        \State leafKvArray $\gets$ scanLeafResult.kvArray
        \State leafKvArraySize $\gets$ scanLeafResult.kvArraySize
        \State continueToNextNode $\gets$ scanLeafResult.continueToNextNode
        \State \Call{sortbykeyasc}{leafKvArray}

        \For{i $\gets$ 0 \textbf{to} leafKvArraySize - 1}
            \State result[resultSize] $\gets$ leafKvArray[i].value
            \State resultSize++
        \EndFor

        \If{continueToNextNode \textbf{AND} leftNode.right $\neq \perp$}
            \State leftNode $\gets$ leftNode.right
        \Else
            \State \textbf{break}
        \EndIf
    \EndWhile

    \State \Call{publishScan}{$\perp$}
    \State \textbf{return} resultSize
\EndFunction
\end{algorithmic}
\label{alg:scan}
\end{algorithm}

\begin{algorithm}
\caption{Fix Underfull Node}
\begin{algorithmic}[1]
\tiny 
\Function{fixUnderfull}{node}
    \If{node = entry \textbf{OR} node = entry.ptrs[0]}
        \State \textbf{return}
    \EndIf

    \While{true}
    \State path $\gets$ \Call{search}{node.searchKey, node}
    \If{path.n $\neq$ node}
        \State \textbf{return}
    \EndIf

    \State right, left $\gets \perp$
    \If{path.nIdx = 0}
        \State sIndex $\gets 1$ \Comment{Sibling is right child}
        \State right $\gets$ parent.ptrs[sIndex]
        \State left $\gets$ path.n
    \Else
        \State sIndex $\gets$ path.nIdx - 1
        \State right $\gets$ path.n
        \State left $\gets$ parent.ptrs[sIndex]
    \EndIf
    \State sibling $\gets$ parent.ptrs[sIndex]

    \State Lock node, sibling, path.p, path.gp
    \If{node.size $\geq$ \text{MIN\_NODE\_SIZE}}
        \State \textbf{return}
    \EndIf
    \If{parent.size $< \text{MIN\_NODE\_SIZE}$ \textbf{OR}
        node, sibling, parent, gParent is marked \textbf{OR}
        node, sibling, parent is TaggedInternal}
        \State Release all locks
        \State \textbf{Continue}
    \EndIf

    \If{node is leaf}
        \If{node.left \textbf{and/or} node.right.parent $\neq$ node.parent}
            \State Lock and check if marked
            \If{marked}
                \State Release all locks
                \State \textbf{Continue} 
            \EndIf
        \EndIf
    \EndIf
    \If{node.size + sibling.size $\leq 2 \times \text{MIN\_NODE\_SIZE}$}
        \State newParent $\gets$ \Call{distributeKeys}{node, sibling, parent} 
        \State gParent.ptrs[path.pIdx] $\gets$ newParent
        \State Mark node, parent, and sibling
        \State Release all locks
        \State \textbf{return}
    \Else
       \State \Call{combineKeys}{left, right, parent, path.pIdx, gParent}
    \EndIf
    \EndWhile
\EndFunction

\end{algorithmic}
\label{alg:fixUnderFull}
\end{algorithm}
\newpage
\begin{algorithm}
\caption{Distribute keys}
\begin{algorithmic}[1]
\tiny 
\Function{distributeKeys}{node, sibling, parent}
\State newNode, newSibling $\gets$ Distribute keys of node and sibling
        \State newParent $\gets$ copy of parent plus pointer to newNode and newSibling, and key between newNode and newSibling
        \If{node and sibling are leafs}
            \State Point newNode and newSibling to each other, using the left and right pointers and connect newNode and newSibling to the leafs linked-list left and right pointers
        \EndIf
        \State \Return newParent
\EndFunction
\end{algorithmic}
\label{alg:distributeLeys}
\end{algorithm}

\begin{algorithm}
\caption{Combine keys}
\begin{algorithmic}[1]
\tiny 
\Function{combineKeys}{left, right, parent, parentIndexInGp, grandParent}
\State newNode $\gets$ Combined keys of right and left nodes
        \If{newNode is leaf}
            \State newNode.left $\gets$ left.left
            \State newNode.right $\gets$ right.right
            \State right.right.left $\gets$ newNode
            \State left.left.right $\gets$ newNode
        \EndIf
        \If{gParent = entry \textbf{AND} parent.size = 2}
            \State entry.ptrs[0] $\gets$ newNode
            \State Mark left, parent, and right
            \State Release all locks
            \State \textbf{return}
        \Else
            \State newParent $\gets$ copy of parent with pointer to newNode instead of left $/$ right
            \State grandParent.ptrs[parentIndexInGp] $\gets$ newParent
            \State Mark left, parent, and right
            \State Release all locks
            \State \Call{fixUnderfull}{newNode}
            \State \Call{fixUnderfull}{newParent}
        \EndIf
\EndFunction
\end{algorithmic}
\label{alg:combineKeys}
\end{algorithm}
\label{FixUnderfull}

\begin{figure}[H]
   \label{fig:MTASetOps} 
    \centering
    
    \begin{subfigure}{1\linewidth}
        \centering
        \includegraphics[scale=0.4]{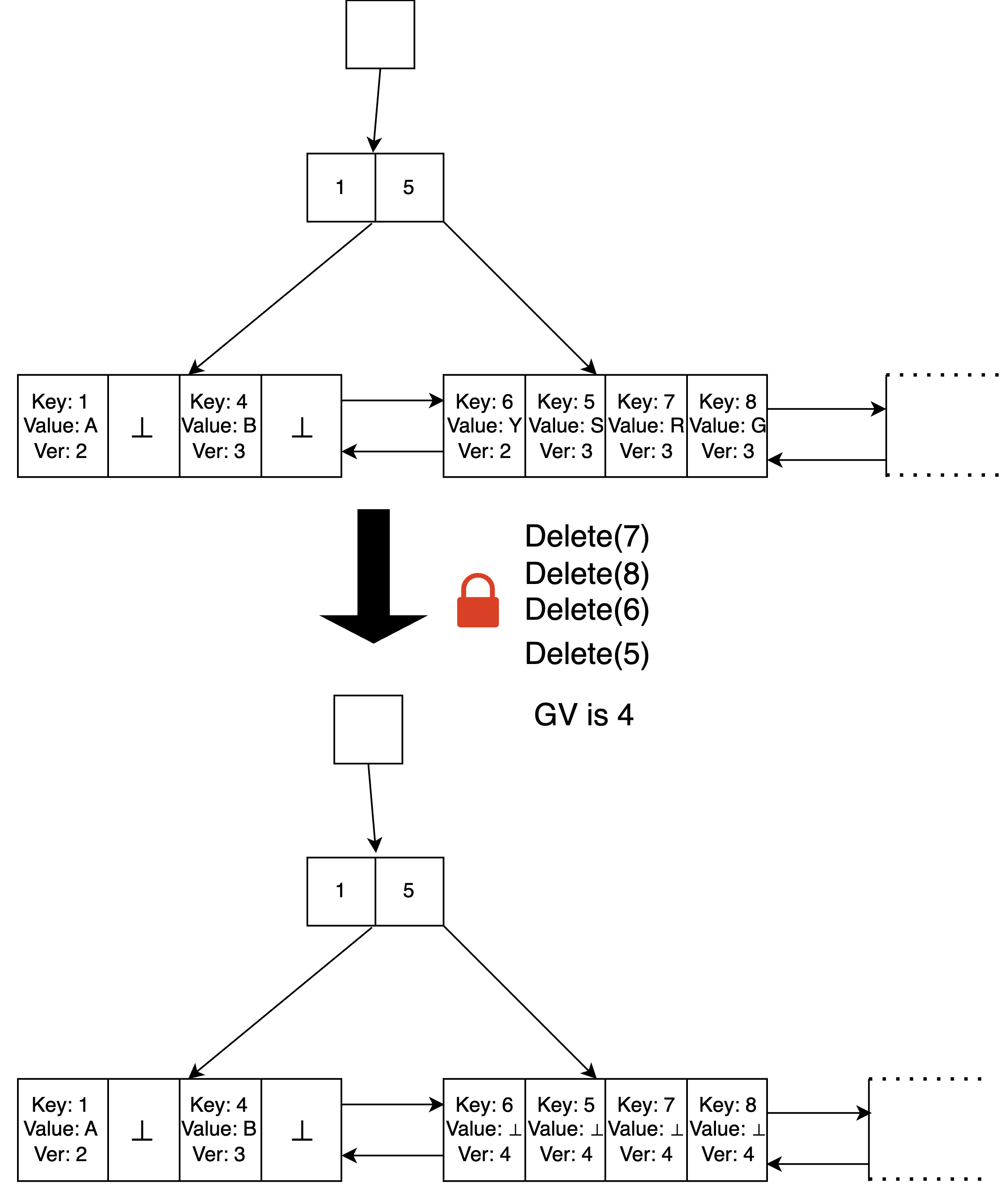}
        \caption{All keys are logically deleted in the right leaf.}
        
        \label{fig:MTASetOp1Delete}  
        \vspace{10px}
    \end{subfigure}
    \begin{subfigure}{1\linewidth}
        \centering
        \includegraphics[scale=0.4]{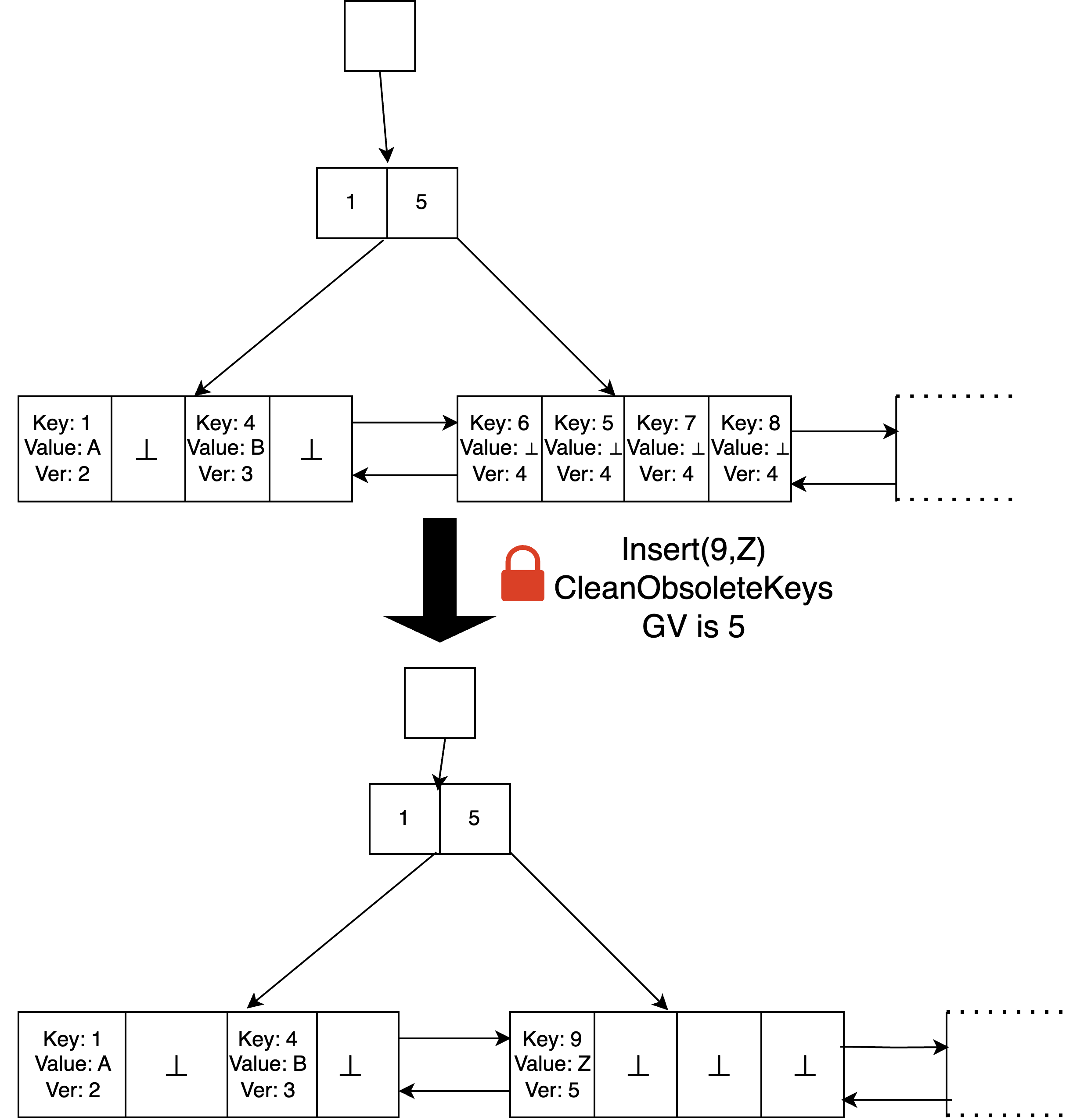}
        \caption{ Insert (9, Z) to trigger the cleanObsoleteKeys function, which will physically remove all keys that have been logically deleted. This is because no keys were logically deleted after all ongoing scans were linearized.}
        \label{fig:MTASetOp2InsertCok}
    \end{subfigure}
\end{figure}

\begin{figure}[H]\ContinuedFloat
    \centering

    \begin{subfigure}{1\linewidth}
        \centering
        \includegraphics[scale=0.4]{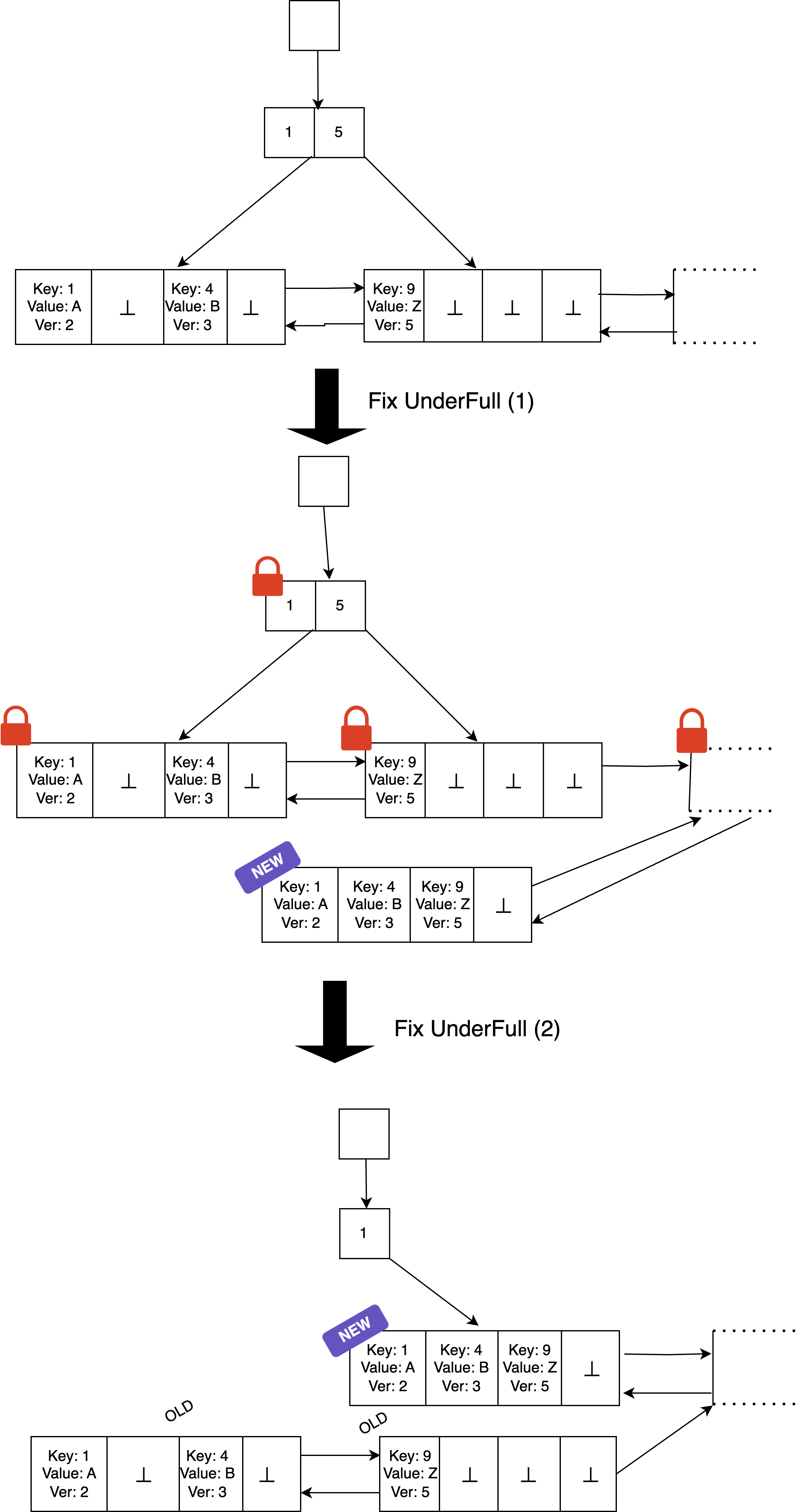}
        \caption{FixUnderfull on an under full leaf node}
        \label{fig:MTASetOp3FixUnderFull}
        \vspace{70px}
    \end{subfigure}

    \begin{subfigure}{1\linewidth}
        \centering
        \includegraphics[scale=0.5]{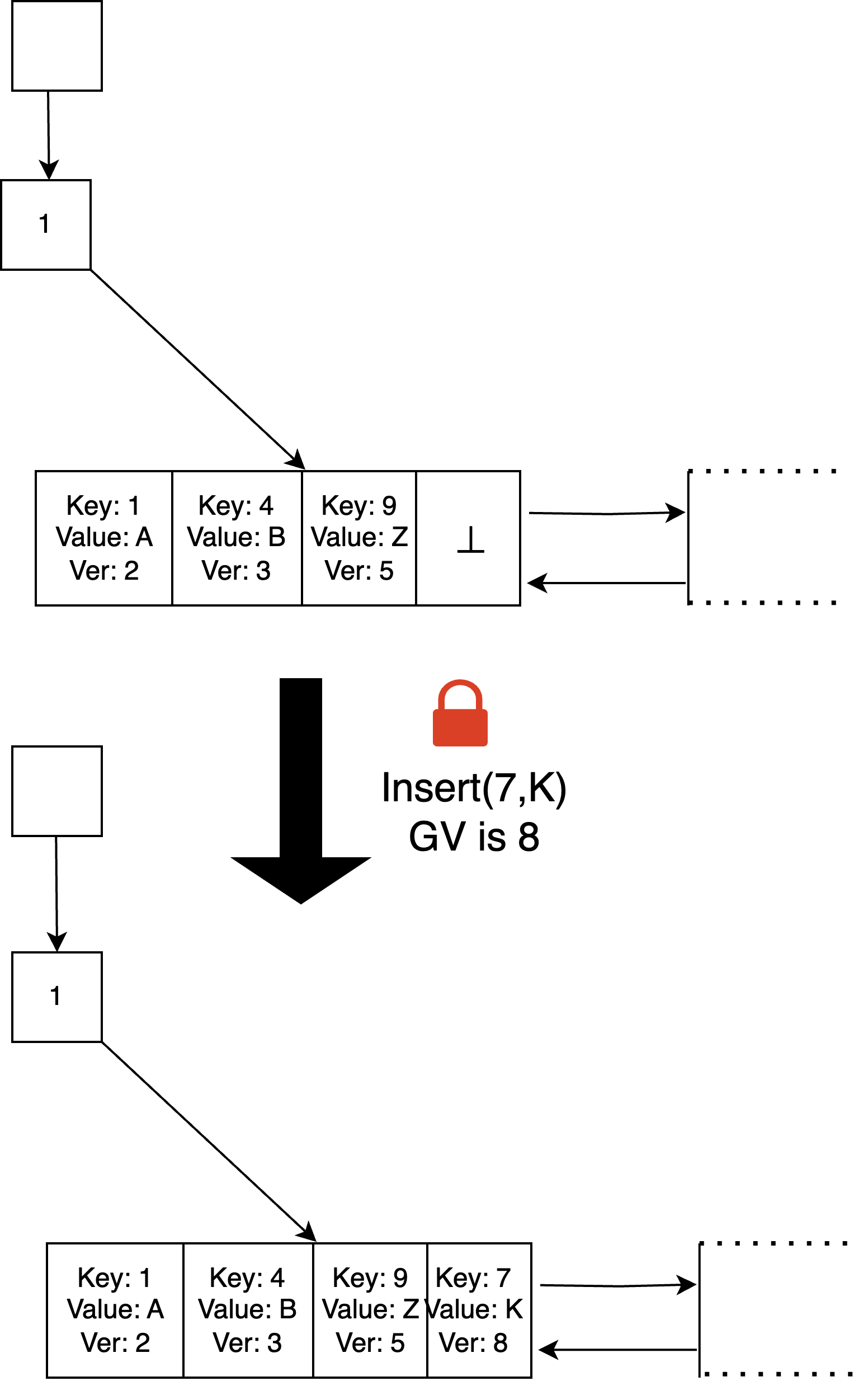}
        \caption{ Insert (7, K), causing the leaf node to become full}
        \label{fig:MTASetOp4Insert}
    \end{subfigure}
    
\end{figure}

\begin{figure}[H]\ContinuedFloat
    \centering

    \begin{subfigure}{1\linewidth}
        \centering
        \includegraphics[scale=0.5]{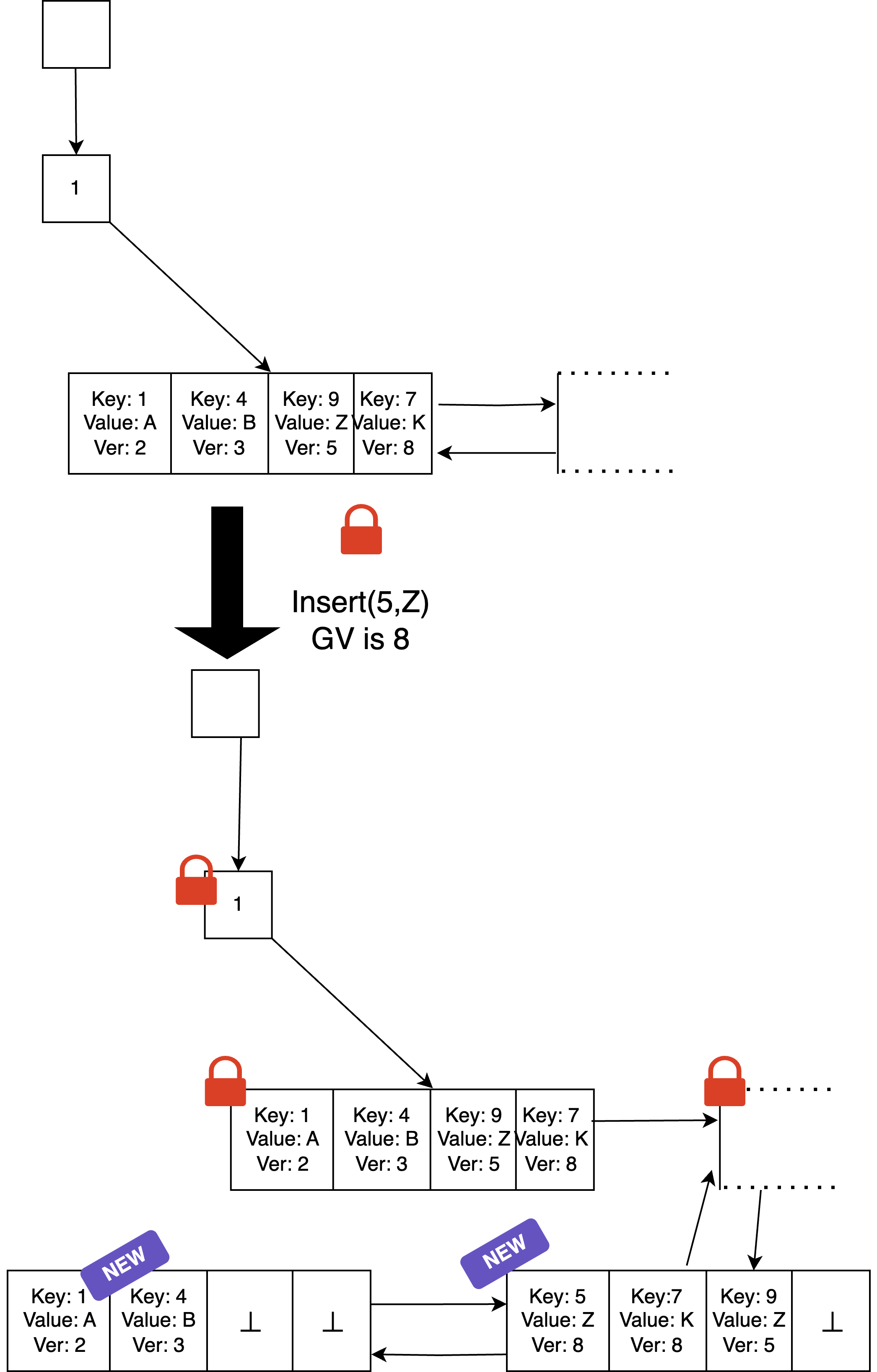}
        \caption{ Insert (5, Z) into the full leaf node, triggering a split insert and creating two leaf nodes.}
        \label{fig:MTASetOp5InsertToFull}
        \vspace{30px}
    \end{subfigure}

    \begin{subfigure}{1\linewidth}
        \centering
        \includegraphics[scale=0.5]{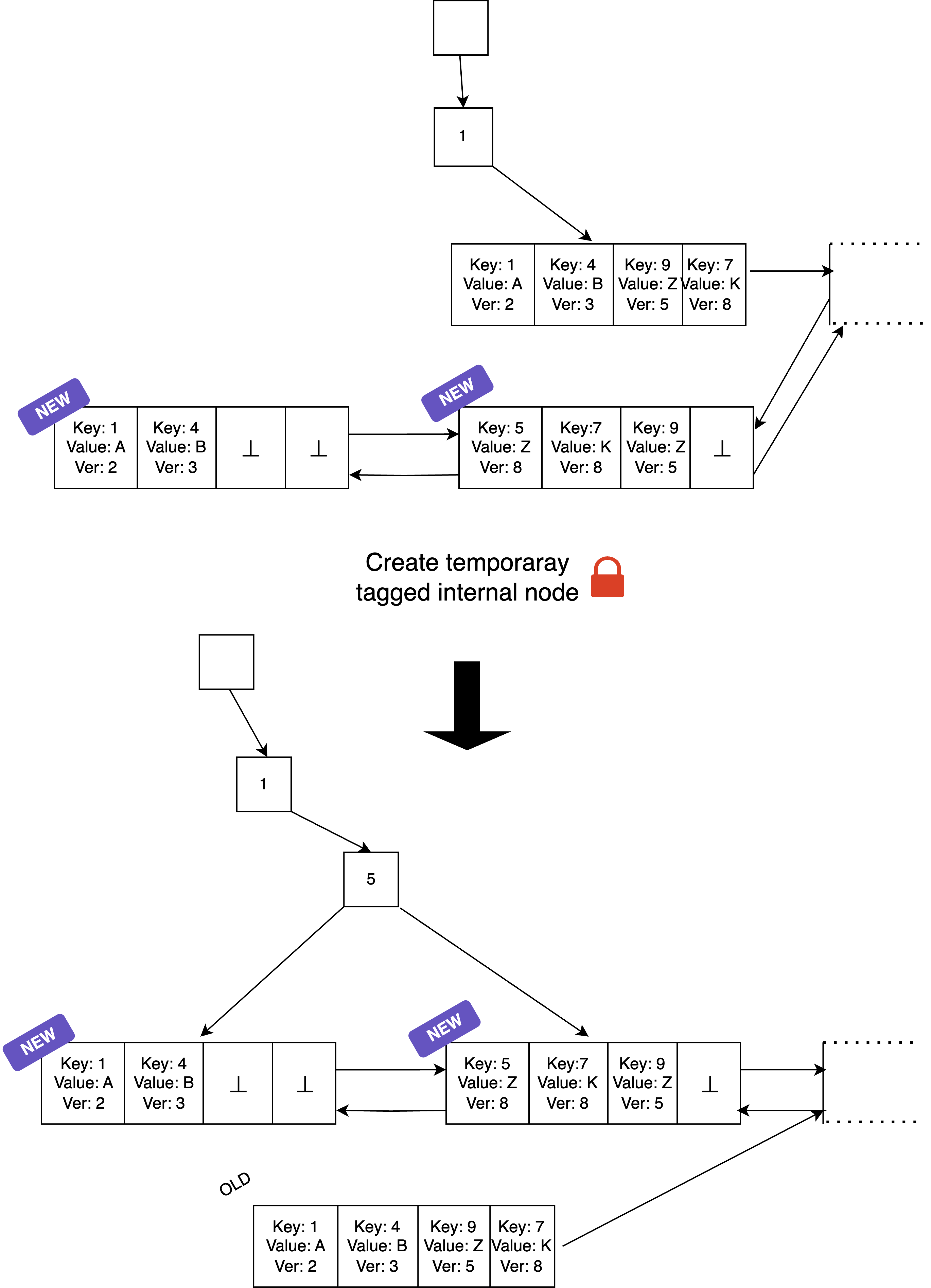}
        \caption{ Create a temporary tagged internal node due to an insertion to a full leaf node.}
       
        \label{fig:MTASetOp6CreateTaggedIndernal}
    \end{subfigure}
\end{figure}

\begin{figure}[H]\ContinuedFloat
    \centering

    \begin{subfigure}{1\linewidth}
        \centering
        \includegraphics[scale=0.5]{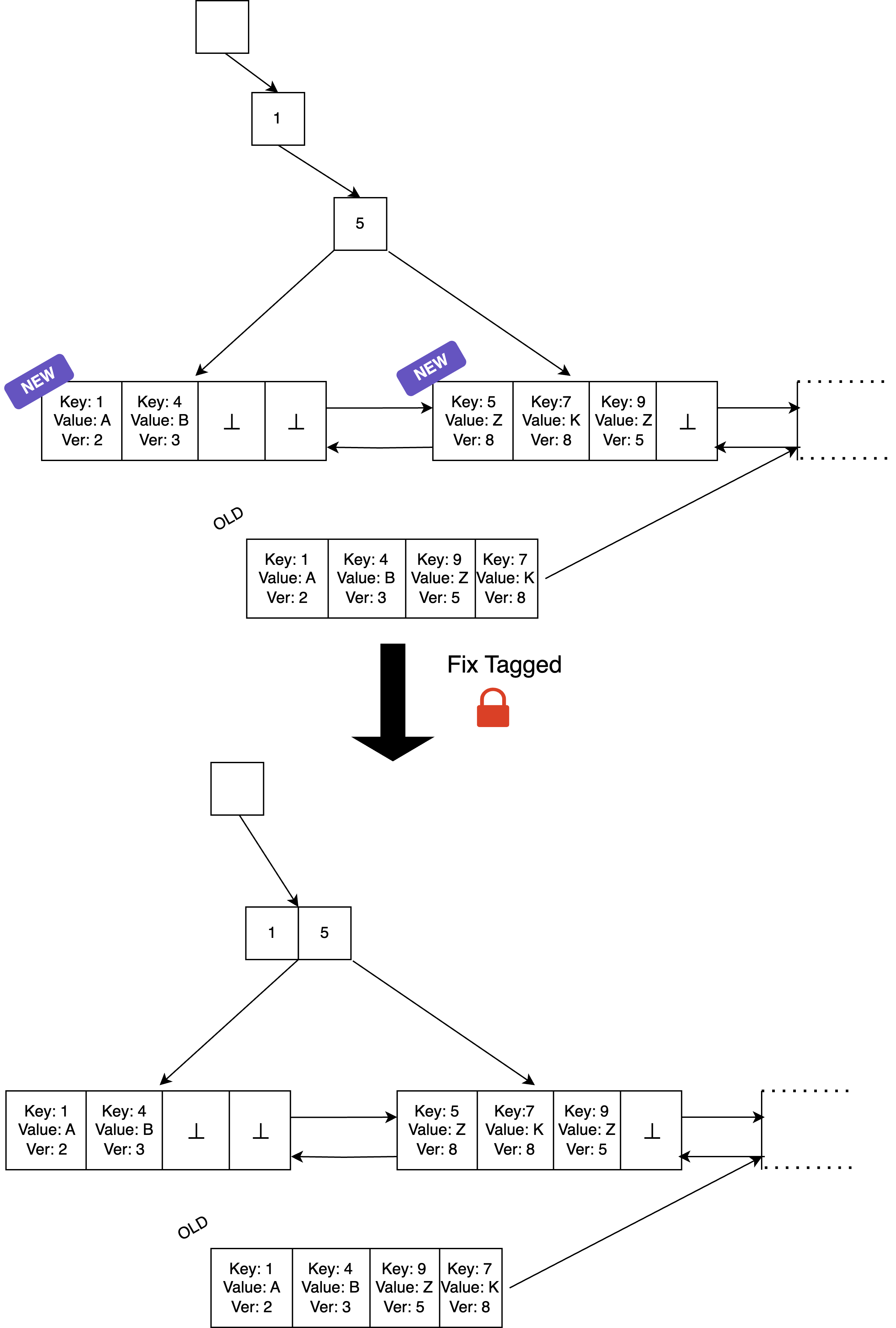}
        \caption{ Invoke FixTagged to address the temporary imbalance by creating a new internal node with keys 1 and 5, and removing the tagged node. In any case, new leaf nodes are linked to the linked list before being linked to the tree, and the old leaf nodes remain linked to the
        linked list after being unlinked from the tree.}
        \label{fig:MTASetOp7FixTagged} 
    \end{subfigure}
    \label{fig:operations}
    \caption{MTASet Operations}
\end{figure}

\begin{figure}
    \centering

    \begin{subfigure}{0.48\textwidth}
        \centering
        \begin{lstlisting}
abstract type Node:
  keys   : K[MAX_SIZE]
  lock   : MCSLock
  size   : int
  marked : bool
        \end{lstlisting}

        \begin{lstlisting}
type Internal extends Node:
  nodes : Node[MAX_SIZE]
        \end{lstlisting}

        \begin{lstlisting}
type Leaf extends Node:
  values  : ValueCell[MAX_SIZE]
  version : int
  left    : Node
  right   : Node
        \end{lstlisting}

    \end{subfigure}%
    \hspace{0.02\textwidth}%
    \begin{subfigure}{0.48\textwidth}
        \centering
        \begin{lstlisting}
type ValueCell:
  value            : V
  version          : int
  previousVersions : BST<int,V> 
        \end{lstlisting}

        \begin{lstlisting}
type PathInfo:
  grandParent   : Node
  parent        : Node
  parentIndex   : int
  node          : Node
  nodeIndex     : int
        \end{lstlisting}

        \begin{lstlisting}
Internal entry // sentinel node
int MAX_SIZE
int MIN_SIZE
int GLOBAL_VERSION
        \end{lstlisting}
    \end{subfigure}

    \caption{MTASet Data Structures}
    \label{fig:mtaset_ds}
\end{figure}

\begin{algorithm}
\tiny
\renewcommand{\alglinenumber}[1]{\tiny#1}
\caption{Search Leaf}
\begin{algorithmic}[1]
\Function{searchLeaf}{key, leaf}
  \State \textbf{RETRY:}
  \State leafVer1 = $leaf$.version
  \State if leafVer1 is odd then go to \textbf{RETRY}
  \State value = $\perp$
  \State \text{Traverse leaf keys array, and if $key$ is found, then value = key’s corresponding value.}
  \State if leafVer1 $\neq$ $leaf$.version then go to \textbf{RETRY}
  \State if {value = $\perp$} then  \Return{(FAILURE, $\perp$)} 
  \State \Return{(SUCCESS, value)}
\EndFunction
\end{algorithmic}
\label{alg:searchLeafShort}
\end{algorithm}

\begin{algorithm}
\caption{Insert}
\label{alg:insertShort}
\begin{algorithmic}[1]
\tiny
\renewcommand{\alglinenumber}[1]{\tiny#1}
\Function{insert}{key, val}
\State \textbf{RETRY:}
\State Search $key$ using the search and searchLeaf algorithms, if found and $val$ is not $\perp$, \Return its corresponding value. // pathInfo object is returned from search in step 2
\State Lock the leaf, and if it is marked, then unlock it and go to \textbf{RETRY}.
\State Verify that $key$ doesn't exist in the leaf and if exists and $val$ is not $\perp$, unlock leaf and \Return its corresponding value.
\If{$key$ is logically deleted and $val$ is not $\perp$ OR $key$ exists and $val$ is $\perp$}
   \State Call updateKeyInIndex(keyIndex,$val$,leaf), unlock leaf, and return $\perp$. 
\EndIf
\If{the leaf is not full}
    \State Call WriteKey($key$, $val$, leaf), unlock leaf, and return $\perp$.
\Else
    \State Call CleanObsoleteKeys(leaf) and if keys were removed, call WriteKey($key$, $val$, leaf), unlock leaf, call FixUnderfull(leaf) and \Return $\perp$.
\EndIf
\State Lock parent node, and if marked, unlock parent, leaf and go to \textbf{RETRY}.
\If{leaf.left or leaf.right are not from the same parent as leaf}
    \State Lock leaf.left or leaf.right.
\EndIf
\If{leaf.left or leaf.right is marked}
    \State Unlock leaf.left or leaf.right, parent, leaf and go to \textbf{RETRY}.
\EndIf
\State $newLeaf \gets$ a taggedInternal node with two children connected to the linked list of leaf nodes and evenly share the content of leaf + $key$ and $val$.
\State $parent.nodes[indexOfLeaf] \gets newLeaf$ // connect newLeaf to the tree
\State Unlock leaf.right or leaf.left (if locked), parent and leaf
\State fixTagged(newLeaf)
\State \Return $\perp$
\EndFunction
\State
\Function{writeKey}{key,value,leaf}
\State Increment the Leaf node version. // odd
\State Write $key$ and $value$ in the first available cells in keys in the values array.
\State Attempt to assign GLOBAL\_VERSION to the value’s version using CAS
\State Increment the Leaf node size 
\State Increment the Leaf node version // even
\EndFunction
\end{algorithmic}
\end{algorithm}

\begin{algorithm}
\tiny
\renewcommand{\alglinenumber}[1]{\tiny#1}
\caption{Delete}
\begin{algorithmic}[1]
\Function{Delete}{key}
    \State \Return \Call{insert}{key, $\perp$}   
\EndFunction
\end{algorithmic}
\label{alg:delete}
\end{algorithm}

\begin{algorithm}
\tiny
\renewcommand{\alglinenumber}[1]{\tiny#1}
\caption{Scan}
\begin{algorithmic}[1]
\Function{Scan}{lowKey,highKey}
\State Publish scan by initializing a cell with scanData at the thread\_id index in the OSA
\State scanVer $\gets$ Fetch-And-Add GLOBAL\_VERSION
\State CAS(scanData.version, $\perp$, scanVer)
\State leftLeaf $\gets$ the leaf that currently contains or is intended to contain $lowKey$, found using the search function
\State continueToNextLeaf $\gets$ true
\While{continueToNextLeaf AND leftLeaf $\neq \perp$}
    \State Traverse the leftNode keys array and collect keys and values where keys are within the [$lowKey$, $highKey$] inclusive range AND values' versions are less or equal to scanVer into the kvArray array.
    \If{a value without a version is encountered}
        \State Read GLOBAL\_VERSION and try to assign it using CAS (Compare-And-Swap)
    \EndIf
    \If{a key greater than $highKey$ is encountered}
        \State continueToNextLeaf $\gets$ false
    \EndIf
    \State Sort the kvArray by key in ascending order
    \State Append kvArray values to result array
    \State leftLeaf $\gets$ leftLeaf.right
\EndWhile
\State Remove scan from OSA by OSA[thread\_id] $\gets \perp$
\State \Return (resultArray, resultArraySize)
\EndFunction
\end{algorithmic}
\label{alg:scanShort}
\end{algorithm}

\section{Correctness}
\label{appendix:correctness}
In this Appendix, we present the correctness properties of MTASet’s operations, focusing on their linearizability. We outline the conditions under which the find, insert, and scan operations maintain linearizability, ensuring that each operation’s result appears as though it were executed instantaneously at a single point in time. For each operation, we discuss how the data structure preserves consistency even amidst concurrent modifications, addressing scenarios such as node unlinking, key versioning, and rebalancing.

\subsection{Linearizability of Find}
The linearizability of the find operation can be reasoned as follows. 
According to invariant \ref{invar_search_key_range_contains_key}, the leaf node at which search(key, target) returns was, at some point, the only leaf node that could potentially contain the key.
The searchLeaf(key, leaf) function will return successfully if, during an interval when the leaf was unlocked, it finds the search key and reads its non-$\perp$ latest value or if it fails to find the key by reading a value whose latest version is $\perp$ or by scanning the entire leaf. Invariant \ref{invar_Key_appear_once} guarantees that the key is unique within the leaf. Since the leaf remains unlocked throughout this interval and nodes are not modified while they are unlocked, the result of the find operation accurately reflects the state of the leaf during that time. If the leaf was part of the tree at any point during this interval, then the find operation can be linearized at that point and considered correct.

However, if the leaf was never part of the tree during the unlocked interval, the find operation linearizes just before the leaf was unlinked. This is because, as stated in invariant \ref{invar_unreached_node_retains}, updates do not alter an unlinked node. Thus, the value returned by find corresponds to what it would be if the find operation had occurred atomically just before the node was unlinked.
To establish this, we need to show that the leaf's unlinking point must have occurred concurrently with the find operation. Using the proof from \cite{srivastava2022elimination}, which proves that each node visited during the search was part of the tree at some point, we can infer that if a node $n$ was not in the tree during an unlocked interval, then the removal of $n$ must have occurred concurrently with the find operation.

\subsection{Linearizability of Insert} \label{correctness:insert}
The linearizability of insert operation in MTASet involves four potential linearization points for an insert(key, val) operation.

An insert operation that successfully locates its target key during the search process follows the exact linearization as a find operation. The return value of the search corresponds to the latest non-$\perp$ value associated with the key (validated by the correctness of find), making it the appropriate value to return for the insert.

Suppose an insert operation finds the key with a non-$\perp$ latest value in leaf $l$ after acquiring $l$’s lock. In that case, it can linearize at any point while holding $l$’s lock. During this lock period, the key cannot be removed from $l$, its associated value remains unchanged, and $l$ cannot be unlinked (as unlinking $l$ necessitates marking it). As the leaf’s version remains even, the associated non-$\perp$ latest value is the correct return value, as per Definition \ref{def_key_in_mtaset}.

An insert operation that inserts a key-value pair into a non-full leaf $l$ or modifies a key whose latest value is $\perp$ (logically deleted) in $l$ linearizes at the second that a version is assigned to the new value that was written to the values arrays. Before this linearization point, the key or its corresponding new value was not present in MTASet since the insert operation accessed $l$ while it was locked without finding the key, or finding a key with a $\perp$ associated value. $l$ remains the only reachable leaf that might contain the key. After the linearization point, as defined in Definition \ref{def_key_in_mtaset}, the key is part of MTASet because it is added to $l$, $l$ remains reachable, and the latest value is non-$\perp$.

For splitting inserts, where searches can detect the change as soon as the pointer to the new subtree is updated in the parent node, the linearization must occur at the write to the parent node. Consider a scenario where a splitting insert writes the new pointer into the parent node $p$ at time $t$. Let $l$ be the leaf that was split and replaced by a tagged node $t$ with children $l1$ and $l2$. Before the write to $p$, the inserted key is not in MTASet since the insert operation reads $l$ while it is locked and does not locate the key with a non-$\perp$ value (and $l$ is the only reachable leaf that might contain the key). However, after the write to $p$, the inserted key is in the tree because it resides in either $l1$ or $l2$, both of which are reachable since $p$ is unmarked and thus accessible (invariant \ref{invar_reachble_is_unmarked}). Other keys in $l$ are unaffected by splitting inserts as they are assigned to either $l1$ or $l2$ during the splitting operation. In cases where the insert operation succeeds, the returned value $\perp$ is correct, given that the insert operation was successful.    
 
\subsection{Linearizability of Scan} \label{correctness:scan}
Scans are linearized when the GV surpasses the version used for collecting values. This typically occurs through Fetch-and-increment and occasionally with the assistance of
cleanObsoleteKeys.
Any key $k$ inserted or deleted and linearized after scan $s$ has been linearized will not be included in $s$ because the version of $k$'s value will be higher than that of $s$.

Due to rebalancing, leaf nodes are continuously and concurrently linked and unlinked to and from the MTASet's underlying tree and the linked list of leaf nodes traversed by the scan. Therefore, we will demonstrate that these modifications do not compromise the accuracy of the values returned by a scan. Let scan(low,high) $s$ be an ongoing scan operation that was linearized at time t, which starts traversing the linked-list of leaf nodes starting from a leaf node $l$, which is designated to hold to smallest key in the range, namely, low, and according to invariant \ref{invar_relxedab}, $l$ is unique. While $s$ traverses the linked list of leaf nodes, the following cases may occur:

The current leaf node was unlinked during a scan visit.
According to Invariant \ref{invar_leaf_node_adj}, the scan can proceed to the next node from an unlinked node. According to Invariant \ref{invar_full_underfull_no_duplicate}, the scan can’t reach the respective node(s), resulting from the unlinked node's rebalance, and, therefore, the scan will not scan a key more than once.

The next leaf node that $s$ will visit was under full and has been replaced. 
A leaf node becomes under full due to keys that were physically removed. Keys are physically removed because they were logically deleted, and all of their previous versions are no longer needed. The scan will visit the respective node(s) resulting from the \texttt{fixUnderFull} procedure, which will miss keys removed before the scan was linearized and would not be collected by the scan anyway.

The next leaf node that $s$ will visit was full and replaced. 
A new key was attempted to be inserted into a full leaf node, but in this case, the full leaf node content is split into two new nodes, which also include the new key. The scan will not collect the new key since it was inserted after the scan was linearized. 

%% file: mtaset.bbl
\begin{thebibliography}{10}
\providecommand{\url}[1]{#1}
\csname url@samestyle\endcsname
\providecommand{\newblock}{\relax}
\providecommand{\bibinfo}[2]{#2}
\providecommand{\BIBentrySTDinterwordspacing}{\spaceskip=0pt\relax}
\providecommand{\BIBentryALTinterwordstretchfactor}{4}
\providecommand{\BIBentryALTinterwordspacing}{\spaceskip=\fontdimen2\font plus
\BIBentryALTinterwordstretchfactor\fontdimen3\font minus \fontdimen4\font\relax}
\providecommand{\BIBforeignlanguage}[2]{{%
\expandafter\ifx\csname l@#1\endcsname\relax
\typeout{** WARNING: IEEEtran.bst: No hyphenation pattern has been}%
\typeout{** loaded for the language `#1'. Using the pattern for}%
\typeout{** the default language instead.}%
\else
\language=\csname l@#1\endcsname
\fi
#2}}
\providecommand{\BIBdecl}{\relax}
\BIBdecl

\bibitem{avni2013leaplist}
H.~Avni, N.~Shavit, and A.~Suissa, ``Leaplist: lessons learned in designing tm-supported range queries,'' in \emph{Proceedings of the 2013 ACM symposium on Principles of distributed computing}, 2013, pp. 299--308.

\bibitem{basin2017kiwi}
D.~Basin, E.~Bortnikov, A.~Braginsky, G.~Golan-Gueta, E.~Hillel, I.~Keidar, and M.~Sulamy, ``Kiwi: A key-value map for scalable real-time analytics,'' in \emph{Proceedings of the 22Nd ACM SIGPLAN Symposium on Principles and Practice of Parallel Programming}, 2017, pp. 357--369.

\bibitem{bronson2010practical}
N.~G. Bronson, J.~Casper, H.~Chafi, and K.~Olukotun, ``A practical concurrent binary search tree,'' \emph{ACM Sigplan Notices}, vol.~45, no.~5, pp. 257--268, 2010.

\bibitem{brown2012range}
T.~Brown and H.~Avni, ``Range queries in non-blocking k-ary search trees,'' in \emph{International Conference On Principles Of Distributed Systems}.\hskip 1em plus 0.5em minus 0.4em\relax Springer, 2012, pp. 31--45.

\bibitem{brown2011non}
T.~Brown and J.~Helga, ``Non-blocking k-ary search trees,'' in \emph{Principles of Distributed Systems: 15th International Conference, OPODIS 2011, Toulouse, France, December 13-16, 2011. Proceedings 15}.\hskip 1em plus 0.5em minus 0.4em\relax Springer, 2011, pp. 207--221.

\bibitem{fomitchev2004lock}
M.~Fomitchev and E.~Ruppert, ``Lock-free linked lists and skip lists,'' in \emph{Proceedings of the twenty-third annual ACM symposium on Principles of distributed computing}, 2004, pp. 50--59.

\bibitem{fraser2004practical}
K.~Fraser, ``Practical lock-freedom,'' University of Cambridge, Computer Laboratory, Tech. Rep., 2004.

\bibitem{kobus2022jiffy}
T.~Kobus, M.~Kokoci{\'n}ski, and P.~T. Wojciechowski, ``Jiffy: A lock-free skip list with batch updates and snapshots,'' in \emph{Proceedings of the 27th ACM SIGPLAN Symposium on Principles and Practice of Parallel Programming}, 2022, pp. 400--415.

\bibitem{braginsky2012lock}
A.~Braginsky and E.~Petrank, ``A lock-free b+ tree,'' in \emph{Proceedings of the twenty-fourth annual ACM symposium on Parallelism in algorithms and architectures}, 2012, pp. 58--67.

\bibitem{natarajan2014fast}
A.~Natarajan and N.~Mittal, ``Fast concurrent lock-free binary search trees,'' in \emph{Proceedings of the 19th ACM SIGPLAN symposium on Principles and practice of parallel programming}, 2014, pp. 317--328.

\bibitem{shafiei2013non}
N.~Shafiei, ``Non-blocking patricia tries with replace operations,'' in \emph{2013 IEEE 33rd International Conference on Distributed Computing Systems}.\hskip 1em plus 0.5em minus 0.4em\relax IEEE, 2013, pp. 216--225.

\bibitem{bernstein1987concurrency}
P.~A. Bernstein, V.~Hadzilacos, N.~Goodman \emph{et~al.}, \emph{Concurrency control and recovery in database systems}.\hskip 1em plus 0.5em minus 0.4em\relax Addison-wesley Reading, 1987, vol. 370.

\bibitem{larsen1995b}
K.~S. Larsen and R.~Fagerberg, ``B-trees with relaxed balance,'' in \emph{Proceedings of 9th International Parallel Processing Symposium}.\hskip 1em plus 0.5em minus 0.4em\relax IEEE, 1995, pp. 196--202.

\bibitem{srivastava2022elimination}
A.~Srivastava and T.~Brown, ``Elimination (a, b)-trees with fast, durable updates,'' in \emph{Proceedings of the 27th ACM SIGPLAN Symposium on Principles and Practice of Parallel Programming}, 2022, pp. 416--430.

\bibitem{black1998dictionary}
P.~E. Black, ``Dictionary of algorithms and data structures,'' 1998.

\bibitem{arbel2018harnessing}
M.~Arbel-Raviv and T.~Brown, ``Harnessing epoch-based reclamation for efficient range queries,'' \emph{ACM SIGPLAN Notices}, vol.~53, no.~1, pp. 14--27, 2018.

\bibitem{lin_cor_cond}
\BIBentryALTinterwordspacing
M.~P. Herlihy and J.~M. Wing, ``Linearizability: a correctness condition for concurrent objects,'' \emph{ACM Trans. Program. Lang. Syst.}, vol.~12, no.~3, p. 463–492, Jul. 1990. [Online]. Available: \url{https://doi.org/10.1145/78969.78972}
\BIBentrySTDinterwordspacing

\bibitem{mtaset}
D.~Manor, ``{MTASet} code,'' \url{https://github.com/danielmanordev/MTASet}.

\bibitem{lea2017concurrent}
D.~Lea, ``Java's {ConcurrentSkipListMap},'' \url{https://docs.oracle.com/javase/8/docs/api/java/util/concurrent/ConcurrentSkipListMap.html}.

\bibitem{fraser2007concurrent}
K.~Fraser and T.~Harris, ``Concurrent programming without locks,'' \emph{ACM Transactions on Computer Systems (TOCS)}, vol.~25, no.~2, pp. 5--es, 2007.

\bibitem{lakshman2016nitro}
S.~Lakshman, S.~Melkote, J.~Liang, and R.~Mayuram, ``Nitro: a fast, scalable in-memory storage engine for nosql global secondary index,'' \emph{Proceedings of the VLDB Endowment}, vol.~9, no.~13, pp. 1413--1424, 2016.

\bibitem{sowell2012minuet}
B.~Sowell, W.~Golab, and M.~A. Shah, ``Minuet: A scalable distributed multiversion b-tree,'' \emph{arXiv preprint arXiv:1205.6699}, 2012.

\bibitem{brown2017techniques}
T.~Brown, \emph{Techniques for constructing efficient lock-free data structures}.\hskip 1em plus 0.5em minus 0.4em\relax University of Toronto (Canada), 2017.

\bibitem{nelson2022bundling}
J.~Nelson-Slivon, A.~Hassan, and R.~Palmieri, ``Bundling linked data structures for linearizable range queries,'' in \emph{Proceedings of the 27th ACM SIGPLAN Symposium on Principles and Practice of Parallel Programming}, 2022, pp. 368--384.

\bibitem{mellor1991algorithms}
J.~M. Mellor-Crummey and M.~L. Scott, ``Algorithms for scalable synchronization on shared-memory multiprocessors,'' \emph{ACM Transactions on Computer Systems (TOCS)}, vol.~9, no.~1, pp. 21--65, 1991.

\bibitem{afek1993atomic}
Y.~Afek, H.~Attiya, D.~Dolev, E.~Gafni, M.~Merritt, and N.~Shavit, ``Atomic snapshots of shared memory,'' \emph{Journal of the ACM (JACM)}, vol.~40, no.~4, pp. 873--890, 1993.

\bibitem{eliminationStack}
\BIBentryALTinterwordspacing
D.~Hendler, N.~Shavit, and L.~Yerushalmi, ``A scalable lock-free stack algorithm,'' in \emph{Proceedings of the Sixteenth Annual ACM Symposium on Parallelism in Algorithms and Architectures}, ser. SPAA '04.\hskip 1em plus 0.5em minus 0.4em\relax New York, NY, USA: Association for Computing Machinery, 2004, p. 206–215. [Online]. Available: \url{https://doi.org/10.1145/1007912.1007944}
\BIBentrySTDinterwordspacing

\bibitem{eliminationArray}
\BIBentryALTinterwordspacing
M.~Moir, D.~Nussbaum, O.~Shalev, and N.~Shavit, ``Using elimination to implement scalable and lock-free fifo queues,'' in \emph{Proceedings of the Seventeenth Annual ACM Symposium on Parallelism in Algorithms and Architectures}, ser. SPAA '05.\hskip 1em plus 0.5em minus 0.4em\relax New York, NY, USA: Association for Computing Machinery, 2005, p. 253–262. [Online]. Available: \url{https://doi.org/10.1145/1073970.1074013}
\BIBentrySTDinterwordspacing

\bibitem{exchanger}
\BIBentryALTinterwordspacing
W.~N. Scherer, D.~Lea, and M.~L. Scott, ``Scalable synchronous queues,'' \emph{Commun. ACM}, vol.~52, no.~5, p. 100–111, may 2009. [Online]. Available: \url{https://doi.org/10.1145/1506409.1506431}
\BIBentrySTDinterwordspacing

\end{thebibliography}
